\documentclass[a4paper,11pt]{article}
\pdfoutput=1 

\usepackage{jcappub,aas_macros} 

\usepackage[T1]{fontenc} 
\usepackage{multicol}

\newcommand{\nsh}[0]{N_{\rm{SHELL}}}
\newcommand{\nsi}[0]{N_{\rm{SIDE}}}
\newcommand{\ndi}[0]{N_{\rm{DIR}}}

\usepackage{hyperref}
\usepackage{xspace}
\usepackage{scalefnt}
\usepackage[squaren]{SIunits}

\addunit{\massh}{\mathit{h}^{-1}M_\odot}
\addunit{\Mpch}{\mathit{h}^{-1}Mpc}
\addunit{\kpch}{\mathit{h}^{-1}kpc}
\addunit{\kms}{km\ s^{-1}}
\addunit{\invMpch}{\mathit{h}\ Mpc^{-1}}

\newcommand\smaller[2][0.85]{{\scalefont{#1}#2}}
\newcommand{\HACC}{\smaller{HACC}\xspace}
\newcommand{\CRKHACC}{\smaller{CRK-HACC}\xspace}
\newcommand{\CAMB}{\smaller{CAMB}\xspace}
\newcommand{\CLASS}{\smaller{CLASS}\xspace}
\newcommand{\REPS}{\smaller{REPS}\xspace}
\newcommand{\SONICC}{\smaller{SONICC}\xspace}
\newcommand{\HEALPIX}{\smaller{HEALPix}\xspace}

\title{\boldmath Improving initialization and evolution accuracy of cosmological neutrino simulations}

\author[a,b,1]{James M. Sullivan,\note{Corresponding author.}}
\author[c]{J.D. Emberson,}
\author[c,d]{Salman Habib,}
\author[c,d]{Nicholas Frontiere}

\affiliation[a]{Department of Astronomy, University of California, Berkeley, CA 94720, USA}
\affiliation[b]{Berkeley Center for Cosmological Physics, University of California, Berkeley, CA 94720, USA}
\affiliation[c]{CPS Division, Argonne National Laboratory, 9700 South Cass Avenue, Lemont, IL 60439, USA}
\affiliation[d]{HEP Division, Argonne National Laboratory, 9700 South Cass Avenue, Lemont, IL 60439, USA}

\emailAdd{jmsullivan@berkeley.edu}
\emailAdd{jemberson@anl.gov}
\emailAdd{habib@anl.gov}
\emailAdd{nfrontiere@anl.gov}

\abstract{
Neutrino mass constraints are a primary focus of current and future large-scale structure (LSS) surveys. 
Non-linear LSS models rely heavily on cosmological simulations -- the impact of massive neutrinos should therefore be included in these simulations in a realistic, computationally tractable, and controlled manner. 
A recent proposal to reduce the related computational cost employs a symmetric neutrino momentum sampling strategy in the initial conditions. We implement a modified version of this strategy into the Hardware/Hybrid Accelerated Cosmology Code (\HACC) and perform convergence tests on its internal parameters. 
We illustrate that this method can impart $\mathcal{O}(1\%)$ numerical artifacts on the total matter field on small scales, similar to previous findings, and present a method to remove these artifacts using Fourier-space filtering of the neutrino density field. 
Moreover, we show that the converged neutrino power spectrum does not follow linear theory predictions on relatively large scales at early times at the $15\%$ level, prompting a more careful study of systematics in particle-based neutrino simulations. 
We also present an improved method for backscaling linear transfer functions for initial conditions in massive neutrino cosmologies that is based on achieving the same relative neutrino growth as computed with Boltzmann solvers.
Our self-consistent backscaling method yields sub-percent accuracy in the total matter growth function.
Comparisons for the non-linear power spectrum with the Mira-Titan emulator at a neutrino mass of $m_{\nu}=\unit{0.15}{\rm eV}$ are in very good agreement with the expected level of errors in the emulator and in the direct N-body simulation.
}

\begin{document}
\maketitle
\flushbottom

\section{Introduction}
\label{sec:intro}

The non-zero mass of neutrinos is perhaps the most well-established manifestation of beyond Standard Model physics \citep{balantekin18, degouvea16, 2019BAAS...51c..64D}. Constraints on total neutrino mass are available from terrestrial experiments with a current upper limit of $0.8$~eV, which is expected to be eventually reduced to a mass sensitivity of $\sim0.2$~eV~\cite{KATRIN2022}. Current cosmological limits from the cosmic microwave background and large-scale structure (LSS) surveys have tighter bounds, at the level of $0.1$~eV~\cite{Planck2020, Alam2021} with some sensitivity to the modeling of systematic effects.
LSS observations play an important role in constraining neutrino masses due to the characteristic suppression of the matter power spectrum, $P_m(k)$, below the neutrino free-streaming scale.
This scale is roughly $k_{fs}(z) \approx (\frac{m_{\nu}}{\mathrm{eV}}) \sqrt{\Omega_{m} a} ~\invMpch$ in matter domination \cite{2006PhR...429..307L}. Since the free-streaming scale is proportional to the neutrino mass, $m_\nu$, and the total amplitude of suppression is proportional to the neutrino energy density, $\Omega_\nu$, observational probes of large-scale structure afford the opportunity, at least in principle, to probe both the individual and total neutrino masses. 

The approach to constraining neutrino mass based on cosmological observations complements particle physics-based probes such as oscillation experiments \citep[e.g.,][]{desalas/etal:2017,esteban/etal:2019} that are sensitive to mass-squared differences between mass eigenstates and tritium $\beta$-decay experiments \cite{robertson/etal:1991,lobashev/etal:1999,kraus/etal:2005,aker/etal:2022} that aim to constrain the absolute neutrino mass scale. Although disentangling individual neutrino masses from cosmological observations remains challenging in practice \citep{archidiacono/etal:2020}, upper bounds placed on the total mass are becoming increasingly precise. For instance, DESI is expected to measure the total mass with an error of $\unit{0.02}{eV}$ \citep{desi:2016} or less when combined with CMB-S4 \cite{2021PhRvD.103b3503X}, which is an order of magnitude improvement over the lower limit of Ref.~\cite{KATRIN2022}.

The uncertainty on measurements of matter two-point statistics in modern surveys is small enough that the effect of massive neutrinos needs to be included for accurate modeling thereof as measured, e.g., by gravitational lensing or spectroscopic galaxy surveys~\cite{2014MNRAS.444.3501B,2015JCAP...11..011P,2016PDU....13...77C,2017MNRAS.470.2617A,2017PhRvD..96l3503V,2018MNRAS.480.5386D,2018PhRvD..98d3526A,2019JCAP...05..041U,2020PhRvD.101h3504I,2020JCAP...04..038P,2022PhRvD.105b3520A}.
The effect of neutrinos on LSS has also been highlighted through alternative probes such as voids \cite{2020JCAP...06..032B,2018JCAP...09..001V,2019JCAP...12..055S, 2019MNRAS.488.4413K, 2021PhRvL.126a1301M}, which have been shown to be particularly sensitive to neutrinos in simulations (however, see Ref.~\cite{2022PhRvD.105l3510B}), or through higher-order statistics like the bispectrum \cite{2020JCAP...03..040H,2018JCAP...03..003R,2019JCAP...05..043C}. 
Accurate modeling of the total matter density field over a wide range of scales is required to exploit these probes.
In particular, it is well-known that neutrinos produce scale-dependent growth of the matter field on scales below their free-streaming scale, which is of order $k_{\mathrm{fs}} \sim \unit{0.1}{\invMpch}$ for a neutrino of mass $m_{\nu} \sim \unit{0.1}{\mathrm{eV}}$ (at $z=0$).
However, the detailed effect of massive neutrinos on the small-scale matter distribution remains less well understood due to non-linear evolution at $k\gtrsim\unit{0.1}{\invMpch}$, especially for statistics beyond the matter power spectrum. While halo model based approaches can be pursued~\cite{abazajian05, Hannestad2020}, more accurate modeling of the matter field on these nonlinear scales is attainable, in principle, through N-body simulations that model the growth of structure in the presence of massive neutrinos.

Cosmological simulations may account for the presence of massive neutrinos at several levels of detail\footnote{The following techniques can be applied to any massive thermal relic species, but we will restrict our attention to massive neutrinos.}. The simplest prescription is to treat massive neutrinos explicitly only at the level of the homogeneous and isotropic background, and including the neutrino perturbations through linear theory, effectively a leading-order expansion in the parameter $f_{\nu}=\Omega_\nu/\Omega_m$ \cite{agarwal/etal:2011,upadhye/etal:2014,maksimova/etal:2021}. This method has been used in the Mira-Titan power spectrum emulator~\cite{2022arXiv220712345M}, to which we will compare our results in Section~\ref{sec:improvements}.
At the next level, massive neutrino perturbations are evolved on a mesh following linear theory and are added as a background component to the gravitational potential during the cold matter (cold dark matter and baryons) force calculation~\cite{brandbyge/hannestad:2009}. This approach is potentially problematic, as the choice to evolve cold matter non-linearly while evolving massive neutrinos at the linear level violates momentum conservation \cite{2014JCAP...11..039B}, in addition to obviously discarding the impacts of non-linear neutrino growth. Various classes of ``linear response'' methods \cite{alihaimoud/bird:2013,liu/etal:2018,chen/etal:2020} take this another step further by allowing the background neutrino mesh to evolve perturbatively in response to the instantaneous non-linear cold matter field of the simulation. Alternative simulation techniques have also been proposed including the computationally ambitious approach of directly solving the Vlasov-Poisson equation in phase space \cite{yoshikawa/etal:2020}, or to solve the neutrino evolution with a Boltzmann hierarchy expansion \cite{banerjee/dalal:2016,dakin/etal:2019,inman/yu:2020}.

Perhaps the most accurate, yet still feasible, choice for non-linear LSS modeling is to treat both cold species and massive neutrinos on equal footing in multi-species particle simulations.
The idea of multi-species simulations, with separate particles for the cold matter and neutrinos, is far from new \cite[e.g.,][]{klypin/etal:1993, gardini98, brandbyge/etal:2008,viel/etal:2010,bird/etal:2012,villaescusanavarro/etal:2014,castorina/etal:2015,inman/etal:2015,emberson/etal:2017,brandbyge/etal:2019,villaescusanavarro/etal:2020,bayer/etal:2021,elbers/etal:2021} and, while superficially straightforward, has a number of subtleties leading to variations in its implementation. A comparison campaign between a number of particle-based approaches as well as some approximate techniques has been recently performed in Ref.~\cite{adamek/etal:2022}. Variations in the particle method typically stem from numerical difficulties associated with modeling fast-moving thermal neutrino particles at early times. One way to overcome this challenge is to adopt a ``hybrid'' approach \cite{brandbyge/hannestad:2010,bird/etal:2018,chen/etal:2022} where the mesh-based strategies described above are utilized at high redshift while switching to particle-based schemes at later times when the neutrino thermal motion has been sufficiently damped. Given the motivation to constrain neutrino masses via cosmological observations, there has also been a recent effort to extend the physical fidelity of neutrino simulations by including baryons as an additional species treated numerically with hydrodynamics \cite{2018ApJ...861...53V,2017MNRAS.471..227M,2020ApJS..249...19R}. 

One of the main challenges posed by neutrino particle simulations is that they are typically limited by shot noise effects due to the high thermal motion of neutrinos at early times. These effects overwhelm the cosmological neutrino signal in the power spectrum unless an extreme number of neutrino particles are simulated \cite{emberson/etal:2017}. To address this issue, Ref.~\cite{banerjee/etal:2018} introduced a method of setting initial conditions for neutrino particles that involves a symmetric sampling of the neutrino momenta across the simulation domain. 
This symmetry drastically reduces the scale-independent shot noise that has plagued earlier simulations, allowing the use of many fewer neutrino particle tracers.
For this reason, a variant of this symmetric scheme (hereafter ``tiling'', to refer to the angular tiling of momenta on the sphere) has been explored by other particle-based neutrino simulations (e.g., Refs.~\cite{brandbyge/etal:2019,villaescusanavarro/etal:2020,bayer/etal:2021}).
However, while tiling succeeds in suppressing neutrino shot noise, it can also introduce numerical artifacts in {\em both} the neutrino and cold matter density fields, as well as their growth, as we will show. 

Another challenge with neutrino simulations relates to how the Newtonian approximation complicates the construction of initial conditions. 
To initialize the cold matter and neutrinos, linear theory transfer functions, generally computed using a Boltzmann solver, must be provided to the simulation as input.
To recover the correct observationally-relevant linear matter power spectrum on large scales at low redshift, one cannot simply provide the transfer functions at the initial redshift of the simulation.
Boltzmann codes (correctly) include non-negligible radiation and metric perturbation terms in the linear equations of motion of the evolved species, which are not included in the Newtonian simulation.
It is possible to include the radiation (and massive neutrino) perturbations explicitly when performing the N-body simulation \cite{2019JCAP...03..022T,adamek/etal:2017}, but usually some form of ``backscaling'' is employed instead \cite{2017JCAP...06..043F}.
For simulations including massive neutrino particles, a linear two-fluid approximation for backscaling transfer functions to account for scale-dependent growth of matter due to neutrinos was developed in Ref.~\cite{zennaro2017}. 
Massive neutrinos, however, do not constitute a fluid.
We address the benefits and shortcomings of treating them as such, and introduce our own improved backscaling scheme.

In this paper, we focus on the two main challenges outlined above. 
First, we present an iterative backscaling method in Section~\ref{sec:backscaling} that is designed to create a self-consistent initial condition framework for massive neutrino cosmologies. Next, we follow the tiling approach~\cite{banerjee/etal:2018}, by implementing it within the Hardware/Hybrid Accelerated Cosmology Code (\HACC; \cite{habib/etal:2016}). We describe our custom implementation of the tiling scheme in Section~\ref{sec:simulations} along with the extra code modifications required to efficiently simulate neutrino particles in a high-performance setting. We present a suite of neutrino simulations in Section~\ref{sec:improvements} that are used to perform a numerical convergence study on the internal parameters of the tiling scheme. We also show results from a mitigation strategy used to remove numerical artifacts arising from the discretization of the initial condition neutrino grid. We compare our final results for the total matter power spectrum with the Mira-Titan emulator and find good agreement. Furthermore, we use our backscaled neutrino transfer functions to show that the simulated neutrino power spectrum systematically deviates from linear theory at early times. We finish with concluding remarks in Section~\ref{sec:conclusions}.

\section{Iterative Backscaling}
\label{sec:backscaling}

Setting up the initial conditions of the cold matter and neutrino particles requires the density and velocity fields at the starting redshift of the simulation, $z_{\rm ini}$. These fields can be obtained in linear theory using a Boltzmann solver such as \CAMB \cite{camb2000} or \CLASS \cite{class2011}. The only caveat is that cosmological simulations typically do not perform the correct linear evolution between the time the species are initialized and the final simulation redshift, $z_{\rm fin}$. In particular, simulations generally employ a Newtonian forward model that omits the radiation and metric perturbations included in the evolution equations of Boltzmann codes. As such, the usual convention is to construct the initial conditions in such a way that the final growth in density perturbations matches the linear theory total matter power spectrum, $P_m(z_{\rm fin},k)$, at the end of the simulation. This process is referred to as ``backscaling''. For $z > z_{\rm fin}$, the simulation and Boltzmann code power spectra will disagree at the linear level, with this disagreement greatest on large scales and early times (see Figure \ref{fig:reps_BS} for an example). In the absense of massive neutrinos, the growth of total matter density perturbations in the Newtonian forward model will be scale-independent meaning that the $z_{\rm fin}$ linear transfer function can simply be rescaled using a multiplicative factor, $D$, known as the growth factor.  

The procedure for computing the scale-independent growth factor in a massless neutrino cosmology involves solving the coupled set of continuity, Euler, and Poisson equations, expressed below at the linear level \cite[e.g.,][]{peebles:1993}:
\begin{align}
\frac{\partial \delta_m}{\partial a} + \frac{\theta_m}{a^2H} &= 0, \\
\frac{\partial \theta_m}{\partial a} + \frac{\theta_m}{a} &= -\frac{1}{a^2H}\nabla^2 \phi, \\
\nabla^2 \phi &= \frac{3}{2}H^2 \Omega_m a^2 \delta_m,
\end{align}
where $\theta \equiv \nabla \cdot {\bf v}_m$ is the velocity divergence field and the subscript $m$ denotes the combined matter field (cold dark matter plus baryons). The solution\footnote{There are two independent solutions to this set of equations but we focus only on the ``growing'' mode while ignoring the ``decaying'' mode that becomes subdominant at late times.} to these equations is found by assuming that $\delta_m$ can be factored into a component that depends only on space and a component that depends only on time. The time-dependent component is known as the growth factor since it describes the growth of density perturbations in time: $\delta_m(a) \propto D(a)$. It follows from the continuity equation that $\theta_m(a) \propto a^2 H(a){\rm d}D/{\rm d}a$. From the above expressions, it is clear that the growth factor depends on the particular cosmology, via the background expansion in the Hubble parameter $H$, as well as through the gravitational source term of the Poisson equation, set by the total matter density, $\Omega_m$. For reasons made clear below, we emphasize the latter dependence by writing the growth factor as $D(a,\Omega_m)$.

Once the growth factor has been computed, backscaling is accomplished by rescaling the transfer function that encodes the time evolution of matter perturbations in linear theory  obtained from the Boltzmann code at the final redshift, $T_m(z_{\rm fin},k)$, in the following manner:
\begin{equation}
T_m^{\rm bs}(z_{\rm ini},k) = \frac{D(z_{\rm ini},\Omega_m)}{D(z_{\rm fin}, \Omega_m)} T_m(z_{\rm fin},k),
\label{eq:tbs}
\end{equation}
where the superscript ``bs'' denotes the backscaled transfer function used as an input to the simulation initial conditions. In the event that cold dark matter and baryon particles are treated as separate species (e.g., in hydrodynamics simulations) then their respective backscaled transfer functions follow as:
\begin{equation}
T_\alpha^{\rm bs}(z_{\rm ini},k) = F_\alpha(z_{\rm ini},k)\frac{D(z_{\rm ini},\Omega_m)}{D(z_{\rm fin}, \Omega_m)} T_m(z_{\rm fin},k),
\label{eq:tbsi}
\end{equation}
where the subscript $\alpha \in \{c,b\}$ denotes cold dark matter and baryons, respectively, and we define the term
\begin{equation}
F_\alpha(z,k) \equiv T_\alpha(z,k)/T_m(z,k).
\end{equation}
It is important that $F_\alpha$ is used in this manner to ensure that each species is initialized with the correct proportion relative to the total matter field at $z_{\rm ini}$ (i.e., backscaling by simply replacing $T_m$ in equation (\ref{eq:tbs}) with $T_\alpha$ would be incorrect due to the fact that the individual solutions for $\delta_{c,b}$ cannot be factored into spatial- and time-independent components as could be done with $\delta_m$). The initial particle velocities are computed by defining an initial velocity transfer function, $T_{\theta,\alpha}$, for each species. This is computed from the continuity equation as the time derivative of the backscaled density transfer function evaluated at the initial time:
\begin{equation}
T_{\theta,\alpha}^{\rm bs}(z_{\rm ini},k) = -a^2H(a) \left.\frac{{\rm d}T_\alpha^{\rm bs}}{{\rm d}a}\right|_{z = z_{\rm ini}}.
\label{eq:tvbs}
\end{equation}
In practice, this can be computed using a finite difference of equation (\ref{eq:tbs}) evaluated at $z_{\rm ini} \pm \epsilon$ \cite{brandbyge/etal:2008,inman/etal:2015} (here we use $\epsilon=0.1$). If only the combined matter field is desired, as is the case in standard single-species simulations, then it follows from equation (\ref{eq:tbs}) that the time derivative of $T_m^{\rm bs}$ in equation (\ref{eq:tvbs}) reduces to the time derivative of $D$, meaning that the velocity transfer function is proportional to the density transfer function. For this reason, single-species simulations generally do not use an explicit velocity transfer function, but rather just numerically evaluate the time derivative of $D$.

In the presence of massive neutrinos, backscaling is complicated by the fact that the total matter field no longer evolves in a scale-independent manner, even in the Newtonian forward model. The reason is related to the large thermal component of neutrino motion that induces a characteristic free-streaming scale, $k_{\rm fs}$. In short, neutrinos cluster like cold matter on scales $k \ll k_{\rm fs}$ but possess sufficient kinetic energy to evade capture in the gravitational potential sourced by cold matter on scales $k \gg k_{\rm fs}$. In other words, neutrinos contribute to the growth of density perturbations only on scales much larger than the free-streaming scale. Hence, the usual backscaling prescription presented above would only be valid on the large scales for which neutrinos contribute to the growth of density perturbations.

In principle, the solution to this problem is to solve the coupled set of growth equations for both the cold matter and neutrinos in order to compute the scale-dependent growth factor that can be used in equation (\ref{eq:tbs}). The difficulty with this approach is that the neutrino Euler equation includes a pressure gradient term that is not trivially expressible in an analytic framework. One strategy is to model neutrinos using the fluid approximation so that a tractable solution can be obtained. This is the approach taken in the code \REPS \cite{zennaro2017} which has attained widespread usage in cosmological simulations \cite{villaescusanavarro/etal:2020,2021ApJ...919...24B,bayer/etal:2021,2018ApJ...861...53V,2018MNRAS.481.2813G,2022PhRvD.105l3510B,2019MNRAS.483..734R,2019JCAP...06..018F,2019JCAP...12..057V,2020JCAP...03..040H,2021JCAP...01..009P,2020MNRAS.499.4905C,adamek/etal:2022}. The drawback of this approach, however, is that the neutrino growth predicted by the fluid approximation can result in $\sim100\%$ discrepancies compared to Boltzmann codes. This feeds back as an error in the total matter growth function (albeit with a much smaller amplitude that scales like $f_\nu$) meaning that the initial conditions will not be seeded in a manner that is consistent with the actual neutrino growth predicted by the Boltzmann solver. Moreover, such large discrepancies in the neutrino growth rate invalidate the use of the fluid approximation in checking whether or not the neutrino particles end up growing correctly in the simulation. 

We propose an alternative strategy that replaces the use of the fluid model with an iterative scheme that assumes the neutrino growth relative to cold matter follows exactly from the output of the Boltzmann code. The basic idea of our method is to replace the value of $\Omega_m$ in the gravitational source term of the Poisson equation with a scale-dependent quantity, $\Omega_m^{\rm eff}(k)$, defining the total amount of matter contributing to the growth of density perturbations on scale $k$. We make use of the known asymptotic limits for growth in massive neutrino cosmologies: on large scales neutrinos contribute to growth so that $\Omega_m^{\rm eff}(k \ll k_{\rm fs}) \rightarrow \Omega_{cb\nu}$ while on small scales neutrinos free-stream meaning that $\Omega_m^{\rm eff}(k \gg k_{\rm fs}) \rightarrow \Omega_{cb}$. On intermediate scales, we should observe a monotonic transition between the asymptotic limits. This is then used to compute a scale-dependent growth factor, $D(z, \Omega_m^{\rm eff})$, that is calculated in the same manner as is commonly done for massless neutrino cosmologies except that the $\Omega_m$ term in the gravitational source term is replaced with $\Omega_m^{\rm eff}(k)$. The determination of $\Omega_m^{\rm eff}(k)$ is performed iteratively until we arrive at a self-consistent solution for all $k$. We circumvent the use of approximate models for neutrino growth by interpolating from the direct output of Boltzmann codes for the neutrino growth relative to cold matter. Of course, this method is still approximate as we collapse the impact of the time-varying neutrino free-streaming length into a single time-integrated quantity $\Omega_m^{\rm eff}(k)$ used to scale the growth between $z_{\rm ini}$ and $z_{\rm fin}$. However, as shown below, the errors introduced in this approximation are small and notably improved compared to the commonly adopted approach. 

The first step in each iteration of our scheme is to set up the initial densities and velocities of the cold dark matter and baryons at the starting time, $z_{\rm ini}$, of the simulation. The densities are set by specifying $\delta_\alpha = T_\alpha^{\rm bs}$ following equation (\ref{eq:tbsi}) with the growth factor evaluated using the scale-dependent $\Omega_m^{\rm eff}(k)$ of that iteration. Note that only the $\Omega_m$ appearing in the gravitational source term in the growth factor is replaced with $\Omega_m^{\rm eff}$ and not any of the background terms in the Hubble factor. The velocities are set by specifying $\theta_\alpha= T_{\theta,\alpha}^{\rm bs}$ following equation (\ref{eq:tvbs}) with $T_\alpha^{\rm bs}$ evaluated at $z_{\rm ini} \pm 0.1$ using the scale-dependent $D(z,\Omega_m^{\rm eff})$. From here, we use a fourth-order Runge-Kutta method to evolve the coupled set of continuity, Euler, and Poisson equations from $a_{\rm ini} = 1/(z_{\rm ini}+1)$ to $a_{\rm fin} = 1/(z_{\rm fin}+1)$ with steps taken in ${\rm ln}a$:
\begin{eqnarray}
\frac{d\delta_{c,b}}{d{\rm ln}a} &=& -\frac{\theta_{c,b}}{aH}, \nonumber \\
\frac{d\theta_{c,b}}{d{\rm ln}a} &=& -\left(\theta_{c,b} + \frac{3}{2}aH\Omega_m\delta_m\right),
\label{eq:fs}
\end{eqnarray}
where $\Omega_m = \Omega_{cb\nu}$ and $\delta_{m} = f_c\delta_c + f_b\delta_b + f_\nu\delta_\nu$ (with $f_\alpha \equiv \Omega_\alpha/\Omega_m$) is the total non-relativistic matter contribution at time $a$. Note that $\Omega_m$ in this expression is fixed as $\Omega_{cb\nu}$ in order to match how the gravitational forces are solved in the simulation. In other words, $\Omega_m^{\rm eff}$ is used {\em only} in the calculation of the growth factor used to set the initial conditions and does not appear anywhere else in the iterative procedure. 

Obviously, evaluating the expression above in a self-consistent manner requires knowledge of $\delta_\nu(a)$. The approach used in the \REPS code of Ref.~\cite{zennaro2017} is to couple the neutrinos in the integration of equation (\ref{eq:fs}) with an additional neutrino pressure term that is modeled with the fluid approximation. Instead of attempting to directly model the growth of neutrino perturbations we rather interpolate from the output of the Boltzmann code. More specifically, we store the ratio $R_\nu = T_\nu/T_{cb}$ from \CAMB at a series of scale factors between $a_{\rm ini}$ and $a_{\rm fin}$ (the ratio is deliberately used -- over interpolating only $T_\nu$ -- so as to cancel out the scale-dependencies associated with the relativistic contributions ignored in the Newtonain forward model). This is then used to set $\delta_\nu(a) = \delta_{cb}(a)R_\nu(a)$ where $\delta_{cb}(a)$ is evaluated at the instantaneous $a$ of the time integrator and $R_\nu(a)$ is linearly interpolated from the sample points. In this way, we avoid the use of any approximations in the modeling of the neutrino pressure term. More importantly, this method is self-consistent since the ultimate goal of the simulation is to model the neutrinos in such a way that their growth relative to the total matter follow the linear theory predictions of the Boltzmann code at early times.

The first iteration in our scheme involves setting $\Omega_m^{\rm eff} = \Omega_{cb\nu}$ for all $k$ in the backscaled growth factor. At the end of the first iteration, we find that the final solution for $\delta_{m}(z_{\rm fin},k)$ matches with that expected from the unmodified \CAMB density transfer function on large scales. On smaller scales, however, we find that $\delta_{m}(z_{\rm fin},k)$ is suppressed since the initial density perturbations were backscaled under the incorrect assumption that neutrinos contribute to growth on those scales. At the end of the first iteration, we mark as sufficiently converged any $k$ for which $\delta_{m}(z_{\rm fin},k)$ is within $\epsilon = 10^{-6}$ of $T_{m}(z_{\rm fin},k)$. In the second iteration, we set $\Omega_m^{\rm eff} = \Omega_{cb}$ for all remaining unconverged scales. The result at $z_{\rm fin}$ is the opposite of the first iteration: on small scales we find that $\delta_{m}$ matches $T_{m}(z_{\rm fin},k)$ while on larger scales $\delta_{m}$ is enhanced with respect to \CAMB since those scales were backscaled using an insufficiently small value for the gravitational source term. We again mark as converged any scales $k$ that are within $\epsilon$ of the target solution. The result after two iterations is that the largest and smallest scales are converged while the results on intermediate scales can be linearly interpolated to make a guess for $\Omega_m^{\rm eff}(k)$ on the next iteration. This procedure is continued with each individual $k$ updating its bracketing bounds of $\Omega_m^{\rm eff}(k)$ around either side of the root of the equation $y = \delta_{m}(z_{\rm fin},k)/T_{m}(z_{\rm fin},k)-1$ until convergence is reached. In practice, we find that four iterations are usually sufficient to satisfy our convergence criteria for $\Omega_m^{\rm eff}(k)$.

\begin{figure}[tbp]
    \center
    \includegraphics[width=0.5\columnwidth]{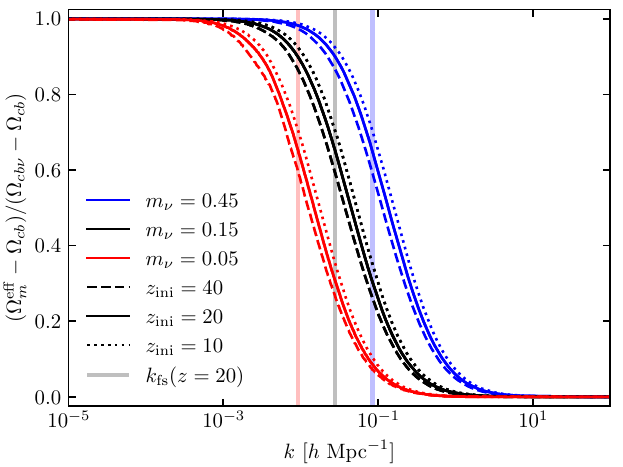}
    \caption{Converged $\Omega_m^{\rm eff}(k)$ for various neutrino masses and initial redshifts (normalized so that $\Omega_{cb\nu}$ is unity and $\Omega_{cb}$ is 0). For reference, the vertical shaded lines show the neutrino free-streaming scale evaluated at $z = 20$ for the corresponding neutrino mass.}
    \label{fig:omegagrav}
\end{figure}
\begin{figure}[tbp]
    \center
    \includegraphics[width=0.7\columnwidth, angle=0]{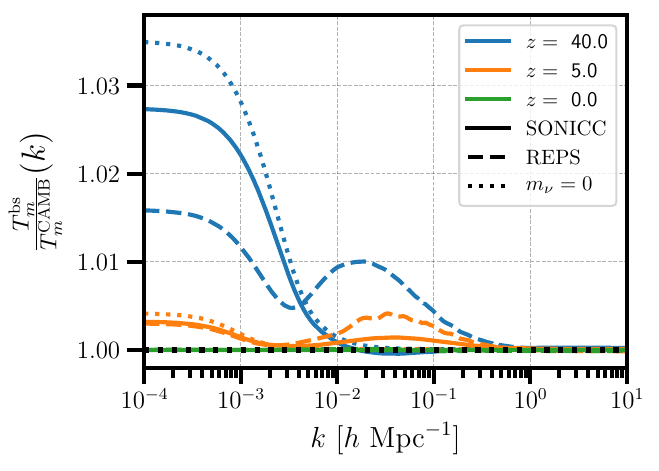}
    \caption{Total matter transfer functions from our iterative backscaling procedure (\SONICC, solid lines) normalized to the unmodified \CAMB transfer functions at various redshifts for a massive neutrino cosmology with $m_\nu = \unit{0.15}{eV}$, $z_{\rm ini} = 40$, and $z_{\rm fin} = 0$. For comparison, the dashed lines show corresponding results from \REPS while the dotted lines show results from a massless neutrino cosmology. For the latter, we use the same cosmological parameters as the massive neutrino case except that we absorb $\Omega_\nu$ into $\Omega_c$ and also increase $N_{\rm eff}$ from 2.046 to 3.046.}
    \label{fig:reps_BS}
\end{figure}
We show in Figure \ref{fig:omegagrav} the converged result for $\Omega_m^{\rm eff}(k)$ for three different neutrino masses ($m_\nu = 0.05$, 0.15, and $\unit{0.45}{eV}$) and three different initial redshifts ($z_{\rm ini}$ = 40, 20, and 10). In each case we set $z_{\rm fin} = 0$ and use the cosmological parameters ($\Omega_c$, $\Omega_b$, $\Omega_\nu$, $\Omega_\Lambda$, $h$, $N_{\rm eff}$) = (0.2684, 0.0491, $m_\nu/(93.14h^2)$, $1-\Omega_{cb\nu}$, 0.6711, 2.046). For each case, we find the expected result that $\Omega_m^{\rm eff} \rightarrow \Omega_{cb\nu}~(\Omega_{cb})$ on large (small) scales with a smooth transition between these limits on intermediate scales. The scales at which $\Omega_m^{\rm eff}$ makes the transition between the two asymptotic limits will roughly span the neutrino free-streaming scales covered between $z_{\rm ini}$ and $z_{\rm fin}$. For reference, the vertical lines in Figure \ref{fig:omegagrav} show $k_{\rm fs}$ evaluated for each neutrino mass at $z = 20$. Note that the departure from the large-scale limit of $\Omega_m^{\rm eff} = \Omega_m$ occurs on larger scales as $z_{\rm ini}$ is increased with fixed neutrino mass. The reason is that $k_{\rm fs} \propto \sqrt{a}$ meaning that the inclusion of earlier times will reflect neutrino suppression on progressively larger scales. Hence, $\Omega_m^{\rm eff}(k)$ depends not only on $m_\nu$, but also on the specific choices of $z_{\rm ini}$ and $z_{\rm fin}$.

In Figure~\ref{fig:reps_BS}, we show the resulting backscaled total matter density transfer function, evaluated using the converged $\Omega_m^{\rm eff}(k)$ in equation (\ref{eq:tbs}), at various redshifts for the $m_\nu = \unit{0.15}{eV}$ case with $z_{\rm ini} = 40$. For reference, we show the backscaled transfer function divided by the unmodified \CAMB transfer function evaluated at each redshift. By construction, our transfer function matches exactly with \CAMB at $z_{\rm fin} = 0$ and deviates at higher redshift. In order to get a rough idea of the level of disagreement we expect with \CAMB at high redshift, we also plot as dotted lines the ratio of backscaled transfer functions to \CAMB for a massless neutrino cosmology. In this case, we use the same cosmological parameters as the massive neutrino cosmology but absorb $\Omega_\nu$ into $\Omega_c$ and set $N_{\rm{eff}} = 3.046$. The backscaled transfer function in this case uses the scale-independent growth factor evaluated with $\Omega_m = \Omega_{cb}$. We do not expect the solid and dotted lines to exactly match, given that they do not correspond to the same cosmology and have differing levels of relativistic neutrino contributions, but we do observe qualitatively similar behavior in terms of how the backscaled transfer functions deviate from \CAMB. In particular, we find that most of the discrepancy appears as an enhancement on large scales at early times. However, we do see slightly different behaviour in the range of scales $10^{-2} \lesssim k/({\rm Mpc}^{-1}h) \lesssim 10^{-1}$ with a small enhancement (suppression) relative to the massless neutrino result at $z = 5$ (40). This difference reflects the approximate nature of integrating the impact of the time-varying neutrino free-streaming scale into the quantity $\Omega_m^{\rm eff}(k)$. Nevertheless, this error appears at only the 
$\sim 0.1\%$ level and is confined to only those scales in the known transition region around $k_{\rm fs}$.

To further illustrate the benefits of our approach, we add dashed lines in Figure~\ref{fig:reps_BS} that correspond to the backscaled transfer functions evaluated using \REPS for the same massive neutrino cosmology\footnote{Following the choices made here, we use their ``Scenario 4'' method that assumes neutrino particles are simulated with a constant mass throughout their evolution.}. It is clear that the backscaled \REPS transfer functions are behaving in a qualitatively different manner than both our iterative backscaling as well as the massless neutrino cosmology. In particular, the \REPS transfer function exhibits $\sim 1\%$ enhancement on scales $10^{-2} \lesssim k/({\rm Mpc}^{-1}h) \lesssim 10^{-1}$ at $z = 40$ and seems to plateau to a different value on large scales than would be expected from comparison with the massless neutrino result. These features arise from the fluid approximation built into \REPS which predicts neutrino growth that deviates from Boltzmann codes at the $\sim100\%$ level (see Figure 3 of \cite{zennaro2017}). This feeds back into the cold matter component, leading to the minor distortions seen in the effective growth factor for the total matter on intermediate scales. An additional drawback from this approach is that the inaccuracies of the fluid model in predicting neutrino growth make it difficult to check if the neutrinos end up growing correctly in the simulation. In this work, we use our backscaled transfer functions to perform a more careful analysis on how well the simulated neutrino growth matches expectations from the Boltzmann code. 

We finish this section by noting that many massive neutrino simulations initialize the cold matter at an earlier redshift, $z_{\rm ini}^{cb}$, than is used for the neutrinos, $z_{\rm ini}^\nu$. In fact, this is the default strategy used by \HACC and employed in the simulations presented in the following sections. In this case, the backscaling method requires an additional modification in order to reflect the fact that the cold matter will grow in a scale-independent manner during the time $z_{\rm ini}^\nu \leq z \leq z_{\rm ini}^{cb}$. The first step is to further backscale the cold dark matter and baryon transfer functions obtained from the iterative method at $z_{\rm ini}^\nu$ to the earlier starting time:
\begin{equation}
T_\alpha^{\rm bs}(z_{\rm ini}^{cb},k) = 
F_\alpha(z_{\rm ini}^{cb},k)
\frac{D(z_{\rm ini}^{cb},\Omega_m)}{D(z_{\rm ini}^\nu,\Omega_m)}
T_{cb}^{\rm bs}(z_{\rm ini}^\nu,k),
\label{eq:tbscb}
\end{equation}
where $T_{cb}^{\rm bs}(z_{\rm ini}^\nu,k)$ is obtained from the iterative backscaling procedure while $F_\alpha(z_{\rm ini}^{cb},k) \equiv T_\alpha(z_{\rm ini}^{cb},k)/T_{cb}(z_{\rm ini}^{cb},k)$ is taken directly from \CAMB. Similarly, for the velocity transfer function we compute:
\begin{equation}
T_{\theta,\alpha}^{\rm bs}(z_{\rm ini}^{cb},k) =
F_{\theta,\alpha}(z_{\rm ini}^{cb},k)
\frac{\dot{D}(z_{\rm ini}^{cb},\Omega_m)}{\dot{D}(z_{\rm ini}^\nu,\Omega_m)}
T_{\theta,cb}^{\rm bs}(z_{\rm ini}^\nu,k),
\label{eq:tvbscb}
\end{equation}
where $T_{\theta,cb}^{\rm bs}(z_{\rm ini}^\nu,k)$ is obtained from the iterative backscaling procedure, the factor $F_{\theta,\alpha}(z_{\rm ini}^{cb},k) \equiv \dot{T}_{\alpha}^{\rm bs}(z_{\rm ini}^{cb},k) / \dot{T}_{cb}^{\rm bs}(z_{\rm ini}^{cb},k)$ with the dots denoting that we apply the finite difference method to the backscaled density transfer functions obtained with equation (\ref{eq:tbscb}), and we define $\dot{D} \equiv a^2H({\rm d}D/{\rm d}a)$. Note that these equations preserve the familiar scale-independent relations $T_{cb} \propto D$ and $T_{\theta,cb} \propto \dot{D}$ seen for the cold matter field in massless neutrino cosmologies. 

It is important to note that the $\Omega_m$ used in equations (\ref{eq:tbscb}) and (\ref{eq:tvbscb}) is set to $\Omega_{cb\nu}$ for all scales even though the neutrinos are not simulated above $z_{\rm ini}^\nu$. In order for the simulation to achieve consistent growth, this convention requires that $\Omega_\nu$ be absorbed into the cold matter particle masses for $z > z_{\rm ini}^\nu$. We have chosen this setup since it guarantees that the cold matter densities and velocities after evolution to $z_{\rm ini}^\nu$ end up matching with the backscaled transfer functions previously computed at $z_{\rm ini}^\nu$, but only on the large scales for which $\Omega_m^{\rm eff} \simeq \Omega_{cb\nu}$. On smaller scales, we incur an inconsistency with the cold matter densities (velocities) being slightly suppressed (enhanced) at the sub-percent level compared to the $T_\alpha^{\rm bs}(z_{\rm ini}^\nu,k)$ $[T_{\theta,\alpha}^{\rm bs}(z_{\rm ini}^\nu,k)]$ previously computed from the iterative method. This is not an issue since the total matter still converges to the same answer at $z_{\rm fin}$, but does mean that we end up with a sub-percent inconsistency in the initial neutrino component as it is set using the $T_\nu^{\rm bs}(z_{\rm ini}^\nu,k)$ that assumed slightly different cold matter density and velocity amplitudes on small scales. An alternative approach would be to set $\Omega_m = \Omega_{cb}$ in equations (\ref{eq:tbscb}) and (\ref{eq:tvbscb}) without changing the simulation particle masses between  $z_{\rm ini}^{cb}$ and $z_{\rm ini}^\nu$, but this would just mean that we incur a minor inconsistency on large scales instead of small scales. In general, starting the cold matter at an earlier time than the neutrinos requires that we incur an inconsistency somewhere due to the discontinuous nature of transitioning from scale-independent to scale-dependent growth at $z_{\rm ini}^\nu$. Pushing this inconsistency to small scales is preferable since these will be the first to undergo non-linear growth that tend to wash out small errors in the initial conditions. Either way, we stress that the inconsistency manifests at the sub-percent level in the neutrino component meaning that its impact on the total matter is exceedingly small.

For convenience, we have created a public code named \SONICC\footnote{\url{https://git.cels.anl.gov/hacc/sonicc}} (Scale-dependent Omega for Neutrino Initial Condition Codes) that uses the iterative backscaling procedure presented above to generate initial density and velocity transfer functions for cold dark matter, baryons, and neutrinos for an arbitrary choice of cosmological parameters as well as starting and stopping redshifts (including the ability to start the cold matter earlier than neutrinos). The code is written in Python3 and uses the public \texttt{camb} module to compute Boltzmann transfer functions at the initial and final times as well as interpolate the relative neutrino growth between these endpoints. It was shown in Ref.~\cite{dakin/etal:2019} that the neutrino transfer functions produced by \CAMB are sensitive to the choice of accuracy settings \texttt{accuracy\_boost} and \texttt{l\_accuracy\_boost}. Our default choice is to take \texttt{accuracy\_boost} = 4 and \texttt{l\_accuracy\_boost} = 5, which means that the neutrino transfer functions are accurate at the few-percent level for $k \gtrsim \unit{0.1}{\invMpch}$.  
With these accuracy settings, the typical runtime of the backscaling code is about 8 minutes, with the bulk of this coming from the computation of \CAMB transfer functions. This can be reduced to under a minute by decreasing the default accuracy settings, but doing so comes at the expense of decreased accuracy in the neutrino transfer functions as well as the resulting $\Omega_m^{\rm eff}(k)$.

\section{Neutrino Initial Conditions and Evolution}
\label{sec:simulations}

Theoretical modeling of the growth of LSS in the presence of massive neutrinos will require an equal or better level of precision as that of observational measurements. Cosmological simulations of structure formation are vital in this regard. In this work, we extend the code \HACC \cite{habib/etal:2016} to evolve neutrinos as an additional N-body particle species alongside the cold matter. \HACC is a cosmological simulation code designed to run performantly with extreme computational loads on all modern supercomputing systems. This performance is inherited here since the inclusion of neutrino particles mainly involves modifications to the initial conditions while preserving the core code structure and algorithmic choices. Recently, a hydrodynamic extension to \HACC, known as \CRKHACC \citep{frontiere/etal:2022}, was developed and we plan to merge the work presented here with the hydrodynamic extension so that fully self-consistent three-species (cold dark matter, baryons, neutrinos) simulations can be performed efficiently at scale.

There are two main challenges associated with the cosmological simulation of neutrino particles: 1) the ambiguities associated with properly backscaling massive neutrino cosmologies in a way that is consistent with the Newtonian forward model; 2) dealing with the high thermal motion that tends to randomize neutrino particles across the simulation domain. To overcome the first issue, we use the iterative backscaling procedure presented above to create more accurate initial transfer functions that are valid for all three non-relativistic matter species. For the second issue, we make use of the ``tiling'' procedure~\cite{banerjee/etal:2018}, which utilizes symmetries in the neutrino thermal momentum sampling in order to significantly reduce the level of shot noise associated with randomization of the neutrino particles. We briefly describe this method below and expand on the minor modifications to its implementation that were used in this work.

In the standard picture, the cosmic neutrino background (CNB) constitutes a thermal relic species from the early universe. More specifically, massive neutrinos decoupled from the early plasma when the universe was roughly one second old and remained free of all interactions save gravity until the present day. During this time, the temperature of the CNB cooled with the expansion of the universe and has a current temperature $T_{\nu,0} \simeq \unit{1.95}{K}$. At early times, the neutrinos behave as a relativistic radiation component while later transitioning into a non-relativistic matter component when the temperature-to-mass ratio, $(k_B T_{\nu,0})/(am_\nu)$, drops below $\mathcal{O}(1)$. The massive neutrino comoving momentum magnitudes, $q$, are set by the relativistic Fermi-Dirac distribution that describes the neutrino phase space after decoupling:
\begin{equation}
    f_{0}(q) = \frac{g_{s}}{2\pi^{2} \hbar^{3}}\frac{q^{2}}{e^{\frac{q}{k_B T_{\nu}}} + 1},
    \label{eqn:f0}
\end{equation}
where $q \equiv a p$, with $p$ the neutrino momentum magnitude and $g_{s}$, the spin factor. 

It is common in cosmological simulations to initialize neutrino particles in an analogous manner to cold matter. First, the neutrino particles are placed on a uniform lattice containing $N_\nu^3$ points with gravitationally-induced displacements and peculiar velocities computed from the initial transfer functions. However, unlike the cold matter, a thermal velocity is also added to each neutrino particle with a random direction vector and random magnitude drawn from the Fermi-Dirac distribution. Occasionally, this stage is modified so that each lattice site initializes a pair of neutrino particles with the pair having random thermal velocities equal in magnitude but opposite in direction \citep{klypin/etal:1993}. In any event, the main drawback of this method is that the thermal velocities are so large that using a random draw results in a density field that is essentially a Poisson sampling of the spatial neutrino distribution, resulting in a shot noise term in the power spectrum that is inversely proportional to the total number of neutrino particles, $P_{\rm shot} \propto N_\nu^{-3}$. Since free-streaming suppresses neutrino perturbations on small scales, a prohibitively large number of particles may be required to resolve the intrinsic neutrino signal below the shot noise level.

In the ``tiling'' method~\cite{banerjee/etal:2018}, the shot noise is reduced by imposing symmetries on the initial neutrino velocity distribution. To begin, this method preserves the use of a neutrino grid containing $N_\nu^3$ points, but associates a total of $\nsh\times\ndi$ particles to each grid point. This collection of neutrino particles is viewed as a set of $\nsh$ independent spherical shells that each contain $\ndi$ particles. The thermal velocity magnitude is varied across the different shells, but held fixed for all particles within the same shell. Hence, only a total of $\nsh$ unique velocity magnitudes are used for the thermal velocities and these are chosen in such a way as to optimize the sampling of the full Fermi-Dirac distribution. Within each shell, the velocity directions are chosen by taking a symmetric discretization of $\ndi$ points on the unit sphere. In the original implementation, these direction vectors are set using \HEALPIX \cite{gorski/etal:2005}, though other variations including minimizing the Coulomb potential of point charges on a sphere \cite{brandbyge/etal:2019} and utilizing a Fibonacci grid \cite{bayer/etal:2021} yield similar results. The crucial aspect of the tiling method is that this discretization of the velocity magnitudes and directions is replicated at each of the $N_\nu^3$ grid points so that a constant flux of neutrino particles pass through adjacent volumes in the simulation domain, thereby suppressing the shot noise.

In the \HACC simulations presented here, our general setup is to initialize a grid of $N_{cb}^3$ cold matter particles with an $N_\nu^3$ neutrino grid that is maximally offset from the cold matter grid in order to reduce the potential impact of artificial particle coupling  \cite{yoshida/etal:2003}. We split the Fermi-Dirac distribution into $\nsh$ bins of equal probability mass and associate the mean value in each bin to the neutrino shells whose limits $\{q^{\rm{min}}_{i},q^{\rm{max}}_{i}\}$ are determined by the inverse Fermi-Dirac cumulative distribution function:
\begin{equation}
    q_{i} = \sqrt{ \frac{\int_{q^{\rm{min}}_{i}}^{q^{\rm{max}}_{i}} dq q^{2} f_{0}(q)}{\int_{q^{\rm{min}}_{i}}^{q^{\rm{max}}_{i}} dq f_{0}(q) } }
    \label{eqn:qbin}
\end{equation}
In the simulations
presented here, we use \HEALPIX to assign the $\ndi$ initial momentum directions. We have also experimented with the use of a Fibonacci grid, but did not observe any significant difference compared to \HEALPIX for a fixed number of directions. The \HEALPIX directions are parameterized by the internal variable $\nsi$ which gives, e.g., $\ndi =$ (12, 48, 192) for $\nsi =$ (1, 2, 4). Motivated by the results of Ref.~\cite{brandbyge/etal:2019}, we modify the original tiling scheme so that the direction vectors are rotated between shells. In particular, for shell $i$, we rotate the $\ndi$ direction vectors by an amount $(i/\nsh)(\pi/2)$ about the symmetry axis of the \HEALPIX discretization (which we align with the $z$ axis). As shown later, this method helps to mitigate numerical effects associated with the tiling scheme symmetries, particularly for smaller $\nsi$. In the recent work of Ref.~\cite{bayer/etal:2021}, the authors advocate for radially displacing each shell from the grid site by an amount proportional to its momentum magnitude in order to reduce a spurious coupling with the cold matter. We do not use this method here (meaning that each shell is sourced at the same grid location), but rather apply a spectral filter in the force solver, discussed below, in order to minimize the back-reaction of the neutrino grid on the cold matter.

The initial displacements and (non-thermal) velocities of the cold matter and neutrino particles are computed using the Zel'dovich approximation (ZA) evaluated on an $N_{cb}^3$ grid with the backscaled transfer functions of each respective species. The same random phases are used for each species with a rotation applied in Fourier space for the neutrino particles to account for their real-space offset from the cold matter grid where the displacement field is sampled. 
We only consider the (unphysical) case of a single massive neutrino here, but plan to accommodate multiple mass eigenstates, and the complexities therein, in future work. 
By default, we initialize the cold matter at $z_{\rm{ini}}=200$ while starting the neutrinos at a lower redshift $z^{\nu}_{\rm{ini}}$ (we consider $z^{\nu}_{\rm{ini}}$ within a range of 10-40). There are competing effects to consider when choosing the starting neutrino redshift, and the values used here are chosen as a compromise between the two limiting regimes. On the one hand, starting neutrinos too early is complicated by the fact that a significant fraction will still be relativistic -- meaning that both their initialization as a non-relativistic species as well as their subsequent evolution will be incorrectly modeled within the Newtonian dynamics of the simulation \cite{nascimento/etal:2021}. On the other hand, starting neutrinos too late means that they become less well-described by linear theory, particularly for the slower-moving population. One way of pushing the starting redshift lower would be with the use of higher-order Lagrangian perturbation theory for initializing both species \cite{2020JCAP...10..034A} (rather than just the ZA, or 1LPT). The impact of starting at lower redshift using higher-order perturbation theory has been studied by several groups \cite[e.g.,][]{michaux/etal:2021,2022MNRAS.516.3821E}; however, a fully self-consistent multi-species approach is not currently available.

The gravitational force calculation in \HACC is decomposed into two parts: (1) the long-range force evaluated using a spectral particle-mesh (PM) method that solves the (filtered) Poisson equation in Fourier space; (2) the short-range force evaluated using direct pairwise interactions with an algorithm optimized for the given hardware (i.e., we employ a particle-particle method on accelerated systems and a tree method on CPU platforms). The two components are delineated by the force-matching scale, $r_s$, that is roughly equal to three times the length of an individual cell in the PM mesh. Our default choice is to use a PM mesh that is of the same size as the initial cold matter particle grid. The main time step is governed by the PM force evaluation while the short-range force is evaluated on a shorter ``subcycled'' time-scale. For the simulations presented here, we take 625 PM steps between $z_{\rm ini}$ and $z_{\rm fin}$ with 4 short-range force evaluations per PM step, following the typical integration procedure used in \HACC gravity-only simulations \citep{habib/etal:2016}. 

The force decomposition and time-stepping remain unchanged with neutrinos except for two minor modifications. Firstly, we have added the flexibility to exclude neutrino particles from the short-range force calculation. This is simply achieved by sorting neutrinos to the end of the particle arrays and only passing the cold matter portion of those arrays to the short-range kernel. This reduces the computational cost of the short-range force calculation, but implicitly assumes that the gravitational impact of neutrinos below the force-matching scale can be safely ignored. We use this method for the simulations presented here, which we justify by the fact that the force-matching scale $r_s \sim \unit{1}{\Mpch}$ while the free-streaming scale $1/k_{\rm fs} \gtrsim \unit{10}{\Mpch}$ for the neutrino mass and box sizes considered here. The second modification is that we have implemented an optional filter that smooths the neutrino density field in Fourier space during the long-range PM force calculation. This is achieved using a simple sharp-$k$ smoothing filter applied in Fourier space via the transformation:
\begin{equation}
\rho_{\nu}^{\rm{sm}}(\mathbf{k}) =
\Theta(k_{\rm cut}-|\mathbf{k}|)\rho_{\nu}(\mathbf{k}),
\label{eqn:filter}
\end{equation}
where $\rho_{\nu}(\mathbf{k})$ is the Fourier transform of the neutrino density field, computed in real space using a cloud-in-cell (CIC) interpolation to the PM mesh, and $\Theta$ is the Heaviside step function with $k_{\rm cut} = \pi N_\nu/L$ chosen to be the the Nyquist frequency, $k_{\rm Nyq}^\nu$, of the neutrino grid. As shown in Section \ref{sec:improvements}, this is done to avoid potential imprinting of the neutrino grid onto the cold matter density field, which we have found to occur if the neutrino mass resolution is sufficiently coarse.  Again, this smoothing assumes that neutrinos are not significantly clustering below the cutoff scale, which we justify since $k_{\rm Nyq}^\nu \gtrsim  10 k_{\rm fs}$ for the main runs presented here. 

The final consideration that we make is in regards to the particle ``overloading'' performed in \HACC. In general, the global simulation volume is subdivided across MPI ranks with each rank assigned a subvolume with side lengths on the order of a few $\unit{10-100}{\Mpch}$ along each axis. This ``alive'' zone of each rank is then extended $\unit{2-10}{\Mpch}$ at the edge of each boundary and filled with particle replicants from the alive zones of neighboring ranks. This is done to ensure proper boundary conditions are utilized in the short-range force for particles at the edge of the alive zone. The overload zone is refreshed at regular intervals (typically every 1-20 PM steps) with the cadence chosen as a balance between minimizing the MPI communication overhead and the propagation of gravitational force errors from the edge of the overload zone inwards to the alive zone (see Ref.~\cite{habib/etal:2016} for more details). 

One of the challenges with neutrino simulations is that the higher momentum neutrinos are still moving relatively fast near the initialization redshift. The problem is that if a neutrino particle is capable of moving further than the width of the overload zone during one refresh cycle then it is possible that all of its replicants end up outside the alive zone at the next refresh. The outcome is that the particle will be removed from the simulation and we may end up carving out neutrinos near the boundaries of each MPI rank. The trivial solution is to increase the refresh rate and/or increase the overload zone. For the simulations presented here, we use a refresh every PM step and increase the overload to $\unit{20}{\Mpch}$ at early times. This choice adds a significant computational overhead and in the future we plan to avoid large overloads by implementing a method to manually pass neutrino particles across MPI ranks when they would otherwise be removed from the simulation. 

\section{Results}
\label{sec:improvements}

We present here results from a suite of \HACC runs that vary the internal parameters of the neutrino tiling scheme as well as the simulation resolution. The full set of runs are enumerated in Table~\ref{tab:sim_table}. In addition to varying the two main parameters of the tiling scheme -- the number of momentum shells $\nsh$ and the number of directions per shell $\ndi$ (as set by the \HEALPIX parameter $\nsi$) -- we also vary the size of the neutrino grid $N_\nu$. This allows us to check for numerical convergence in the power spectra as one parameter is varied while the others are held fixed. The main set of runs use a simulation box of width $L = \unit{250}{\Mpch}$ containing $N_{cb}^3 = 512^3$ cold matter particles though we also consider two other runs with a larger box of $L = \unit{1000}{\Mpch}$. We also explore the impact of the neutrino starting redshift by varying $z_{\rm{ini}}^{\nu} = (10,20,40)$ with $z_{\rm ini}^{cb}$ held fixed at 200. 

In Section~\ref{subsec:convergence} we present a numerical convergence study for the neutrino tiling parameters. We then show in Section~\ref{subsec:spikes} that one challenge with the tiling method is that the Fourier imprint of the neutrino grid can transfer onto the cold matter power spectrum if the simulation resolution is sufficiently coarse. We proceed in Section~\ref{subsec:emulator} with a comparison of the simulated total matter power spectrum to emulator predictions~\cite{2022arXiv220712345M} and finish in Section~\ref{subsec:linear} with a closer look at the linear theory growth of neutrino power. In all of the further analysis, we examine the power spectra of the cold matter and neutrino components which are computed using a CIC interpolation of each species onto a mesh containing $512^3$ cells. 

\begin{table}[t!]
    \centering
    \begin{tabular}{|c|c|c|c|c|c|c|c|c|c|}
    \hline
    Name & $N_{cb}$ & $N_{\nu}$ & $\nsh$ & $\ndi$ & $z^{\nu}_{\rm{ini}}$ & Filter &  $L$ & Rotate $\phi$ \\
    \hline  
    \texttt{fid\_nophi} & 512 & 128 & 5 & 12  & 20 & Yes & 250 & F \\
    \texttt{fid\_nofil} & 512 & 128 & 5 & 12  & 20 & No & 250 & T \\
    \texttt{fid} & 512 & 128 & 5 & 12  & 20 & Yes & 250 & T\\
    \texttt{med\_N$\nu$}& 512  & 256 & 5 & 12 & 20 & Yes & 250 & T \\
    \texttt{hi\_N$\nu$}& 512  & 512 & 5 & 12  & 20 & No & 250 & T \\
    \texttt{med\_NDIR}& 512  & 128 & 5 & 48  & 20 & Yes & 250 & T \\
    \texttt{hi\_NDIR}& 512  & 128 & 5 & 192  & 20 & Yes & 250 & T \\
    \texttt{med\_NSH}& 512  & 128 & 10 & 192 & 20 & Yes & 250 & T \\
    \texttt{hi\_NSH}& 512  & 128 & 20 & 192 & 20 & Yes & 250 & T \\
    \texttt{coarse\_hi\_NDIR}& 512  & 64 & 5 & 192  & 20 & Yes & 250 & T \\
    \texttt{lo\_z} & 512  & 128 & 5 & 192 &  10 & Yes & 250 & T \\
    \texttt{hi\_z} & 512  & 128 & 5 & 192  & 40 & Yes & 250 & T \\
    \texttt{lb\_loN\_nofil} & 256  & 64 & 5 & 12  & 40 & No & 1000 & T \\
    \texttt{lb\_loN} & 256  & 64 & 5 & 12  & 40 & Yes & 1000 & T \\
    \hline
    \end{tabular}
    \caption{Simulation suite considered in this work. All simulations initialize the cold matter at $z_{\rm ini}^{cb} = 200$ in a box of side length $L$ while the neutrinos are initialized later, at $z_{\rm ini}^\nu$. In each case, we use the cosmological parameters, ($\Omega_c$, $\Omega_b$, $\Omega_\nu$, $\Omega_\Lambda$, $\sigma_8$, $h$, $N_{\rm eff}$) = (0.2684, 0.0491, $m_\nu/(93.14h^2)$, 1-$\Omega_{cb\nu}$, 0.8, 0.6711, 2.046), for a single massive neutrino species with $m_\nu = \unit{0.15}{eV}$. The cold matter contains $N_{cb}^3$ particles while the total number of neutrino particles is given by the product $N_{\nu}^{3} \times \nsh \times \ndi$. In each run, the long-range force is calculated using a PM mesh containing $N_g = N_{cb}$ cells per side; the short-range force is evaluated only on the cold matter and uses a Plummer softening length of $0.1L/N_g$. The ``Filter'' column denotes whether the sharp-$k$ force filter in equation (\ref{eqn:filter}) is applied to the neutrino density field during the long-range force evaluation. The ``Rotate $\phi$'' column indicates whether the momentum shells were individually rotated, as described in Section \ref{sec:simulations}.
    }
    \label{tab:sim_table}
\end{table}

\subsection{Convergence Study of Neutrino Initialization Parameters}
\label{subsec:convergence}

\begin{figure}[t!]
    \center
    \includegraphics[width=6in, angle=0]{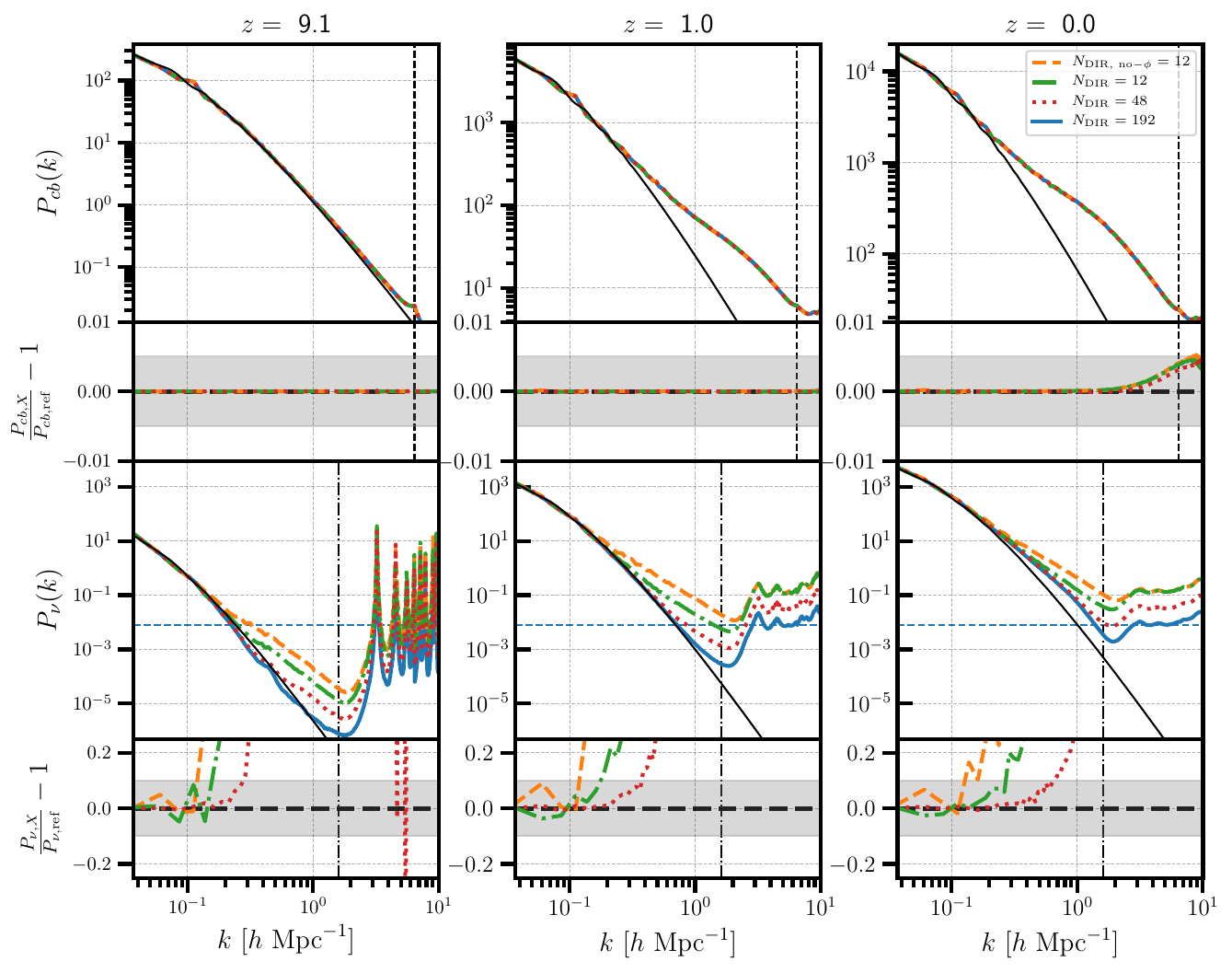}
    \caption{Cold matter (upper two rows) and neutrino (lower two rows) power spectra for different choices of number of neutrino particle velocity directions (\texttt{fid} [green dash-dotted], \texttt{med\_NDIR} [red dotted], \texttt{hi\_NDIR} [blue solid]), as well as without rotating \HEALPIX momentum shells (\texttt{fid\_nophi} [orange dashed])).
    The ratio of the power spectrum at each choice of parameters relative to the power spectrum of the \texttt{hi\_NDIR} run ($\ndi = 192$) is shown in the residual panels (second and fourth rows).
    Shaded bands in the residual panels are at $0.5\%$ for cold species and $10\%$ for neutrinos.
    Plots are shown at several redshifts in the different columns.
    Backscaled linear theory is also shown for reference as the thin black curves.
    In the third row of panels, the dashed blue line corresponds to the expected Poisson shot noise for the \texttt{hi\_NDIR} run which contains the largest neutrino particle count. 
    The vertical dashed black line in the upper panels is the Nyquist mode of the initial cold species grid, while the dash-dotted black line in the lower panels is the same for the initial neutrino grid.
    }
    \label{fig:tiling_num}
\end{figure}

We begin by varying $\ndi$ with all other parameters held fixed. In Figure \ref{fig:tiling_num} we show the cold matter and neutrino power spectra at redshifts $z = 9$, 1, and 0 for the \texttt{fid\_nophi}, \texttt{fid}, \texttt{med\_NDIR}, and \texttt{hi\_NDIR} runs. All runs contain $N_{cb}^3 = 512^3$ cold matter particles with an $N_\nu^3 = 128^3$ neutrino particle grid consisting of $\nsh = 5$ momentum shells. In addition, all of the runs except for \texttt{fid\_nophi} use the method described in Section \ref{sec:simulations} of rotating the direction vectors of each shell. Comparing the dashed orange and dot-dashed
green lines in Figure \ref{fig:tiling_num} show that this rotation has a significant impact on the neutrino power spectrum, both at early and late times. This behavior was observed previously~\cite{brandbyge/etal:2019} and likely reflects the fact that the rotation reduces the correlation in neutrino particle trajectories of adjacent momentum shells. Though not shown here, we find that the relative impact of the rotation decreases as we increase $\ndi$. This can be attributed to the fact that the direction vectors become increasingly isotropic as the discretization is made increasingly fine.

Comparing the dot-dashed green, dotted red, and solid blue lines in Figure~\ref{fig:tiling_num} shows that the neutrino power spectrum is quite sensitive to the number of direction vectors. At $z = 9$, the $\ndi = 12$ and 48 runs are converged to the $\ndi = 192$ run at the $10\%$ level up to $k \simeq 0.1$ and $\unit{0.3}{\invMpch}$, respectively. This improves at $z = 0$ with the $10\%$ agreement extending to $k \simeq 0.3$ and $\unit{0.7}{\invMpch}$ for the two runs. 
Hence, the $\ndi = 192$ run appears to be reasonably converged (at the $\sim10\%$ level) up to half the Nyquist frequency of the neutrino grid at $z = 0$, though more directions may be required to reach a similar level of convergence at higher redshift. 
Note, however, that as we increase $\ndi$, the neutrino power spectrum on scales $k \gtrsim \unit{0.1}{\invMpch}$ converges to a result at $z=9$ that is markedly below the linear theory curve; we will discuss this discrepancy in Section~\ref{subsec:linear}. 
The large differences in neutrino power spectra for $k > k_{\rm Nyq}^\nu$ simply reflect different levels in the effective neutrino shot noise of each run. The shot noise level is governed by the total particle count, as shown by the horizontal line tracing $P_{\rm shot} = L^3/(N_\nu^3\nsh\ndi)$ for the \texttt{hi\_NDIR} run. This is also the level at which the neutrino power spectrum saturates to in standard simulations that draw random velocity magnitudes and directions. The fact that we are able to resolve the neutrino power spectrum below $P_{\rm shot}$ for $k \leq k_{\rm Nyq}^\nu$ thus confirms the utility of the tiling method, as also verified in Refs.~\cite{banerjee/etal:2018,bayer/etal:2021,brandbyge/etal:2019}. Note that the neutrino power spectrum still saturates to $P_{\rm shot}$ for $k > k_{\rm Nyq}^\nu$ with the presence of noticeable spikes observed at the Fourier modes of the neutrino grid. We will explore this topic in more detail in Section~\ref{subsec:spikes}.

Despite the large variance observed in neutrino power spectra with varying $\ndi$, we see that the cold matter power spectra are relatively unchanged between each run. This is particularly true at high redshift and linear scales though we do observe a $\sim0.5\%$ scatter near the cold matter Nyquist frequency at $z = 0$. As further shown below, we generally find that the neutrino power spectra exhibit systematic numerical convergence as one parameter is varied while all others are held fixed. The cold matter power spectra, on the other hand, mainly show small random noise on non-linear scales. We attribute this to the fact that minor changes in the neutrino distribution are not strong enough to significantly alter the cold matter on linear scales (since $f_\nu$ is small), but are able to seed noise, particularly in the shot-noise dominated regime, that becomes amplified with non-linear growth. Given that the noise is random in nature (i.e., it fluctuates between runs in a non-systematic manner) and confined to highly non-linear scales, its presence does not indicate that the tiling scheme fails to numerically converge, rather that care must be taken to understand the amplitude of random error in the total matter distribution on small scales. 

\begin{figure}[t!]
    \center
    \includegraphics[width=6in, angle=0]{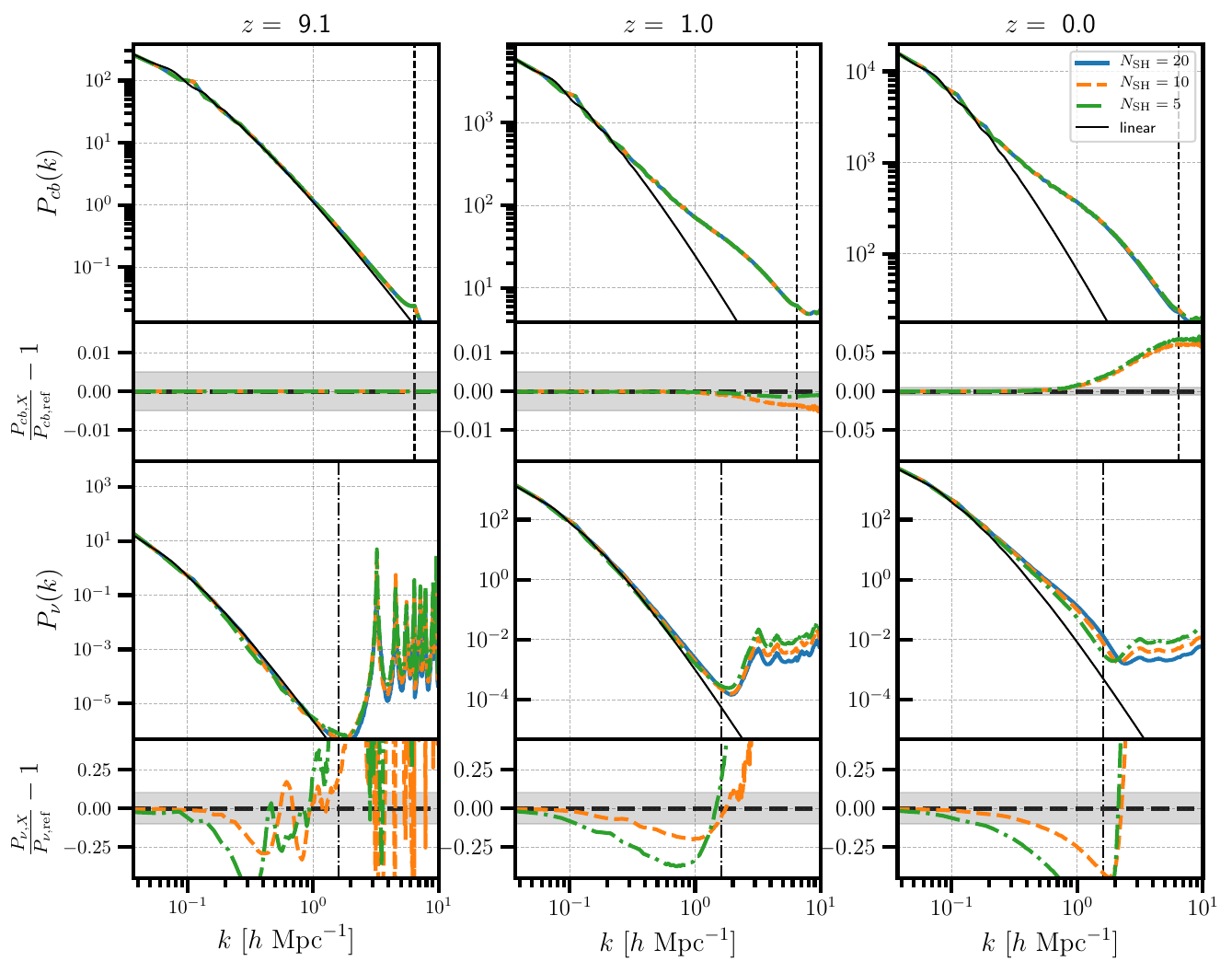}
    \caption{ 
    Cold matter (upper two rows) and neutrino (lower two rows) power spectra for different choices of number of neutrino momentum bins (\texttt{fid} [green dash-dotted], \texttt{med\_NSH} [orange dashed], \texttt{hi\_NSH} [blue solid]).
    All simulations have $\ndi = 192$.
    The ratio of the power spectra with respect to the \texttt{hi\_NSH} run with $\nsh = 20$ is shown in the residual panels (second and fourth rows).
    Shaded bands in the residual panels are at $0.5\%$ for cold species and $10\%$ for neutrinos.
    Plots are shown at several redshifts in the different columns.
    The vertical dashed black line in the upper panels is the Nyquist mode of the initial cold species grid, while the dash-dotted black line in the lower panels is the same for the initial neutrino grid.
    Backscaled linear theory is also shown for reference as the thin black curves.
    }
    \label{fig:tiling_sh}
\end{figure}

Next we consider the case where we vary $\nsh$ while keeping the other parameters fixed.
Similar to Figure~\ref{fig:tiling_num}, Figure~\ref{fig:tiling_sh} shows cold matter and neutrino power spectra at redshifts $z=9, 1, ~ \mathrm{and}~ 0$ for the \texttt{fid}, \texttt{med\_NSH}, and \texttt{hi\_NSH} runs, which have $\nsh=5,10,20$ respectively (all at $\ndi=192$\footnote{We also found qualitatively similar results when varying $\nsh$ when the number of directions was fixed to $\ndi=12$.}).
On the largest scales we find good agreement, with all three runs following linear theory closely.
On intermediate scales, the neutrino power in the three runs starts to diverge at around $k\approx 0.2~\invMpch$ at the $10\%$ level.
As expected, the $\nsh=10$ neutrino power agrees more with that of the $\nsh=20$ run than the $\nsh=5$ run does, and this is true at all redshifts.
For $k>k_{\rm Nyq}^{\nu}$, the difference between the neutrino power spectra is due to a change in shot noise since the $\nsh=20$ (10) run has four times (twice) as many neutrino particles as the $\nsh=5$ run.
Meanwhile, the cold species power is essentially unaffected by the change in $\nsh$ at the $0.1\%$ level for linear scales. At later times, we again see fluctuations in the cold matter power on the non-linear scales near the Nyquist mode. The random nature of this noise is evident in the flipping and reordering of the green and orange curves when comparing the $z = 1$ and $z = 0$ residual panels.
Overall, we find a less dramatic effect on the power spectra of both species when varying the number of shells compared to the number of directions; in agreement with the initial exploration of Ref.~\cite{banerjee/etal:2018}.

\begin{figure}[t!]
    \center
    \includegraphics[width=6in, angle=0]{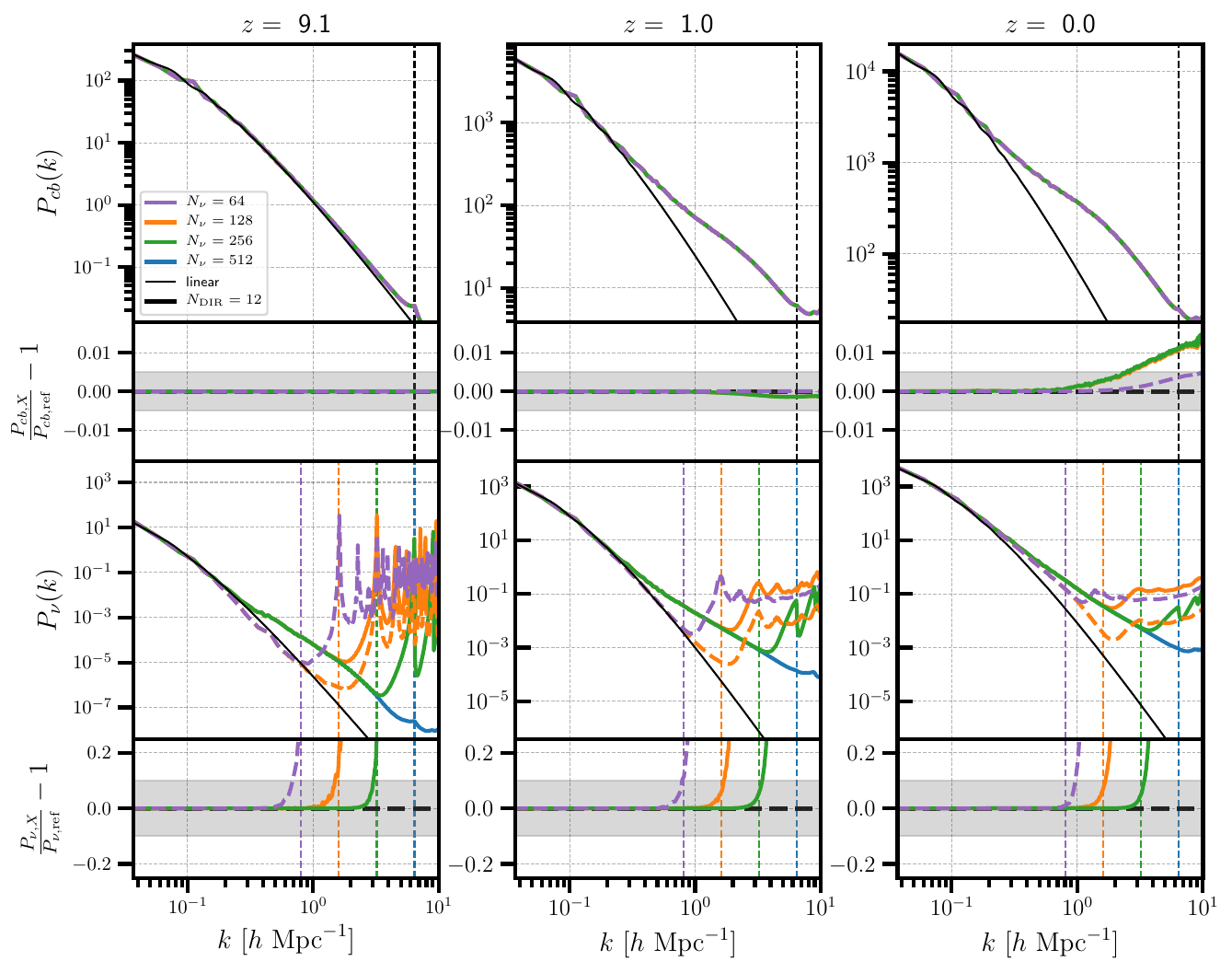}
    \caption{
    Cold matter (upper two rows) and neutrino (lower two rows) power spectra while varying the resolution of the neutrino particle grid, $N_\nu$, for runs with $\ndi=12$
    (solid lines) and $\ndi=192$ (dashed lines).
    Residuals are shown with respect to the $N_{\nu}=512$ run (blue solid line, \texttt{hi\_N$\nu$})
    for the $\ndi=12$ runs, and with respect to the $N_{\nu}=128$ run (orange dashed line, \texttt{coarse\_hi\_NDIR}) for the $\ndi=192$ runs (i.e., they are made with respect to the largest value of $N_\nu$ for each set of $\ndi$ runs).
    Plots are shown at several redshifts in the different columns.
    Shaded bands in the residual panels are at $0.5\%$ for cold species and $10\%$ for neutrinos.
    The vertical dashed black line in the upper panels is the Nyquist mode of the initial cold species grid, while in the lower panels, the dotted colored lines correspond to the Nyquist modes of the corresponding neutrino grid resolutions (with colors indicated in the legend).
    Backscaled linear theory is also shown for reference as the thin black curves.
    }
    \label{fig:tiling_conv}
\end{figure}

Having characterized how the cold species and neutrino power spectra change with $\ndi$ and $\nsh$, we now consider varying the resolution of the neutrino grid $N_{\nu}$ in Figure~\ref{fig:tiling_conv}.
Following our earlier convention, we show the cold matter and neutrino power spectra for several values of $N_{\nu}$ in different colors, where solid (dashed) lines correspond to runs with $\ndi=12$ (192) and $\nsh=5$ held fixed in each case.
Since all simulations have $N_{cb} = 512^{3}$, we can consider this analysis to represent making different choices of the ratio $N_{\nu}/N_{cb}$.
In terms of the neutrino power spectra, the general trend of increasing $N_\nu/N_{cb}$ is independent of the value of $\ndi$. Namely, we find that the neutrino power spectrum is insensitive to the choice of $N_\nu/N_{cb}$ up until the Nyquist frequency of the neutrino particle grid. This is easily seen in the bottom panel, where each line traces a residual of zero up until the vertical dotted line indicating its Nyquist wavenumber. As mentioned before, comparing the two sets of $\ndi$ runs show that the neutrino power spectrum is significantly enhanced on intermediate scales for $\ndi=12$. 
Furthermore, Figure~\ref{fig:tiling_conv} shows that up until a comoving wavenumber of $k\approx 3.0 ~\invMpch$ (at $z=0$), the $N_{\nu}=256$ and $N_{\nu}=512$ neutrino power spectra agree at the 10\% percent level. This comparison is important for considering the computational cost of the tiling method, as the former run has a factor of 8 fewer neutrino particles than the latter. We see the same general behavior in the cold matter power as was shown in the previous two tests. Namely, the cold matter power displays random noise on small non-linear scales. As before, the cold matter is relatively insensitive to changes in $N_\nu/N_{cb}$ on linear scales.

The results presented above show that numerical convergence at the $\sim10\%$ level in the neutrino power spectra requires $\ndi \gtrsim$ 100 direction vectors and $\nsh \gtrsim 10$ momentum shells. The cold matter power is not significantly impacted by choices in these parameters on linear scales, but does exhibit random noise at the $\sim1\%$ level on non-linear scales at late times. We suspect that this error is seeded by shot noise from the neutrinos on small scales (which scales inversely with the total number of neutrino particles) and becomes amplified by non-linear growth. The main challenge with the tiling method is that the requirement $\ndi \gtrsim$ 100 and $\nsh \gtrsim 10$ implies that $N_\nu/N_{cb} \lesssim 1/10$ for the number of neutrino particles to not significantly exceed the cold matter count. The coarseness of the neutrino particle grid limits the smallest scale at which neutrinos are effectively resolved and, as shown below, requires extra care to prevent discreteness effects from artificially back-reacting on the cold matter.

\subsection{Suppression of Neutrino Grid Artifacts}
\label{subsec:spikes}

The runs shown in Figure \ref{fig:tiling_conv} display characteristic small-scale spikes in the neutrino power spectra with locations that depend on the value of $N_\nu$. These spikes were also observed in Ref.~\cite{brandbyge/etal:2019} and correspond to resonances in the neutrino particle grid which are picked up as essentially delta functions in Fourier space. More specifically, these spikes occur at multiples of the Nyquist frequency of the neutrino particle grid (e.g., the first two spikes occur at $2k_{\rm Nyq}^\nu$ and $2\sqrt{2}k_{\rm Nyq}^\nu$). Of course, this is not a feature unique to the neutrino particles; if we were to measure the cold matter power spectrum on a fine enough mesh then we would likewise observe a series of spikes at the Fourier modes corresponding to the cold matter particle grid~\cite{joyce/marcos:2007}. In other words, single-species N-body simulations also contain the Fourier imprint of the initial conditions grid that are subsequently captured in the force solver provided the force resolution is finer than the particle grid separation. These grid features are generally considered to insignificantly impact evolution in the single-species case due to the efficient gravitational transfer of power from large to small scales \cite[e.g.,][]{joyce/etal:2009}. The result is that gravity tends to broaden the spikes in Fourier space until the gravitational growth on small scales surpasses the shot noise level, erasing the spikes altogether. This gravitational broadening can be seen in the lower panels of Figure~\ref{fig:tiling_conv} for the neutrinos and is also examined more extensively in Ref.~\cite{brandbyge/etal:2019}. However, it is not clear what impact these features may have in multi-species simulations especially if the Nyquist frequencies of the particle grids are mismatched, as is the general case here.

\begin{figure}[t!]
    \center
    \includegraphics[width=2.75in, angle=0]{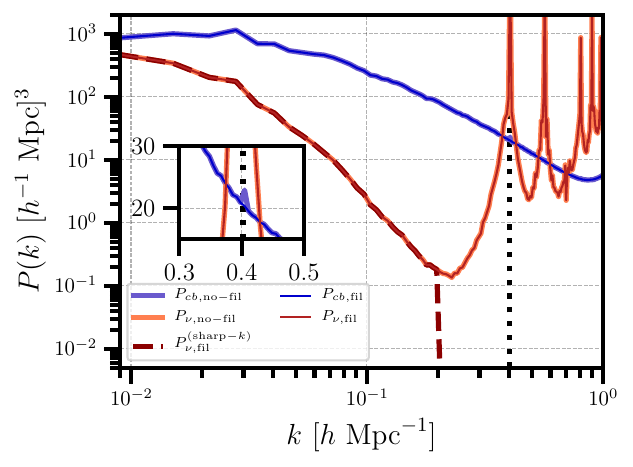}
    \includegraphics[width=3in, angle=0]{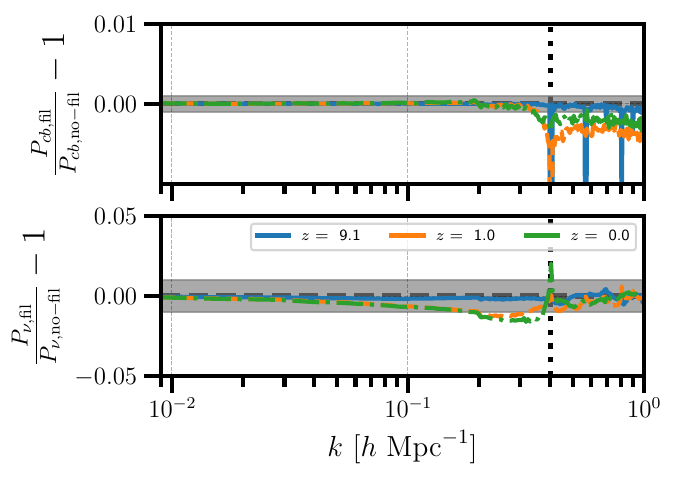}
    \caption{ \textit{Left}: Cold species (blue) and neutrino (red) power spectra for the \texttt{lb\_loN\_nofil} (thick shaded lines) and \texttt{lb\_loN} (thin solid lines) runs at $z=5$. The vertical dotted black line denotes twice the neutrino particle grid Nyquist frequency which corresponds to a prominent spike in the neutrino power spectrum as well as the cold species power spectrum in the non-filtered case (see the inset for a zoom-in on the cold species power spectra). The dashed red line shows the neutrino power spectrum when the sharp-$k$ filter is applied. \textit{Right}: The ratio of the cold species (top) and neutrino (bottom) power spectra from these two runs at redshifts $z=9.1, 1.0, 0.0$.
    Shaded regions are at $\pm 0.1\%$ for cold species and $\pm 1\%$ for neutrinos.}
    \label{fig:spike_ev}
\end{figure}

To study this issue more closely, we ran two simulations with $N_\nu < N_{cb}$ and varying mass resolutions to test whether we could clearly identify the imprint of the neutrino grid on the cold matter. Indeed, we found that for sufficiently coarse resolutions, it is possible for the neutrino grid to transfer its Fourier signal onto the cold matter power spectrum. This is demonstrated in Figure \ref{fig:spike_ev} where we show the power spectrum at $z = 5$ for the \texttt{lb\_loN\_nofil} simulation that contains $N_{cb}^3 = 256^3$ cold matter particles with an $N_\nu^3 = 64^3$ neutrino particle grid in an $L = \unit{1000}{\Mpch}$ box. First, we see the characteristic spikes in the neutrino power spectrum which extend prominently above the neutrino shot noise level. Furthermore, we also clearly observe a small spike in the cold matter power spectrum situated at the location of the first neutrino spike, as shown in the plot inset. The presence of this spike unambiguously demonstrates that the numerical imprint of the neutrino particle grid is being transferred onto the cold matter density field via the force solver. We checked for similar spikes in the cold matter power spectra of two other simulations, \texttt{fid\_nofil} and \texttt{fid}, but did not observe any, indicating that this imprinting depends on the simulation resolution. We suspect that the probability of the neutrino grid imprinting on the cold matter field is enhanced when the neutrino shot noise is more comparable to the amplitude of the cold matter power spectrum at early times. In other words, care must be taken to control for this issue if the neutrino mass resolution -- which, like the shot noise, scales as $L^3/(N_\nu^3\nsh\ndi)$ -- is too coarse.

Of course, the absence of spikes in the cold matter power spectrum of the other two simulations does not mean that the cold matter was not still influenced by the neutrino shot noise level on small scales. This is the motivation behind the use of the sharp-$k$ filter described in Section \ref{sec:simulations} that forcefully removes any contributions from the neutrino grid in the long-range force solver. For reference, the dark red dashed line in Figure \ref{fig:spike_ev} shows the neutrino power spectrum measured when this filter is applied. The abrupt truncation of the neutrino power spectrum for $k > k_{\rm Nyq}^\nu$ means that the numerical signal of the neutrino grid is completely hidden from the long-range force calculation. To test the impact of this filter on the final result, the right panels of Figure \ref{fig:spike_ev} compare $P_{cb}$ and $P_\nu$ from the original non-filtered run, \texttt{lb\_loN\_nofil}, with its filtered counterpart, \texttt{lb\_loN}. 
Here we observe that the cold matter power spectrum of the filtered run is essentially unchanged below $2 k_{\rm Nyq}$, the location of the first neutrino grid spike. At higher wavenumbers, we see sharp spikes occurring at the Fourier modes of the neutrino grid which correspond to the erasing of these signatures from the cold matter field in the filtered run. At lower redshifts, gravitational broadening of the spikes in the non-filtered run dissipates their signal on small scales.
In the lower panel showing the neutrino residuals, we see that the filtering leads to only precent-level changes in the evolution of the neutrino density field. In other words, the filtering strategy effectively removes clear numerical artifacts from the neutrino grid without strongly impacting the small-scale growth of cold matter or neutrino density perturbations.

We reiterate that the sharp-$k$ filter is only justified if neutrinos are not contributing to the growth of density perturbations beyond $k_{\rm Nyq}^\nu$. This assumption becomes increasingly inaccurate the closer $k_{\rm Nyq}^\nu$ gets to the neutrino free-streaming scale, and therefore should only be applied if the resolution is such that $k_{\rm Nyq}^\nu \gg k_{\rm fs}$. This condition can be somewhat alleviated if the filter is turned off at lower redshift when both the free-streaming length is smaller and the intrinsic neutrino power has grown closer or above the shot noise level. We note that Ref.~\cite{bayer/etal:2021} uses an alternative method to mitigate impacts associated with the neutrino grid which involves radially displacing each neutrino shell during the initialization by an amount proportional to its momentum magnitude. We tried this method but did not find it to be effective for the \texttt{lb\_loN\_nofil} run presented above. In any event, we consider the sharp-$k$ filter to be a more direct method for preventing an artificial back-reaction from the discretization of the neutrino grid. We also tested other functional forms for the filter, but found that the smoothing required to adequately remove the neutrino spikes using obvious candidates (Gaussian and Hann functions) was so broad that the filter artificially suppressed cold matter growth on larger scales.

\subsection{Emulator Comparison}
\label{subsec:emulator}

\begin{figure}[t!]
    \center
    \includegraphics[width=5.5in]{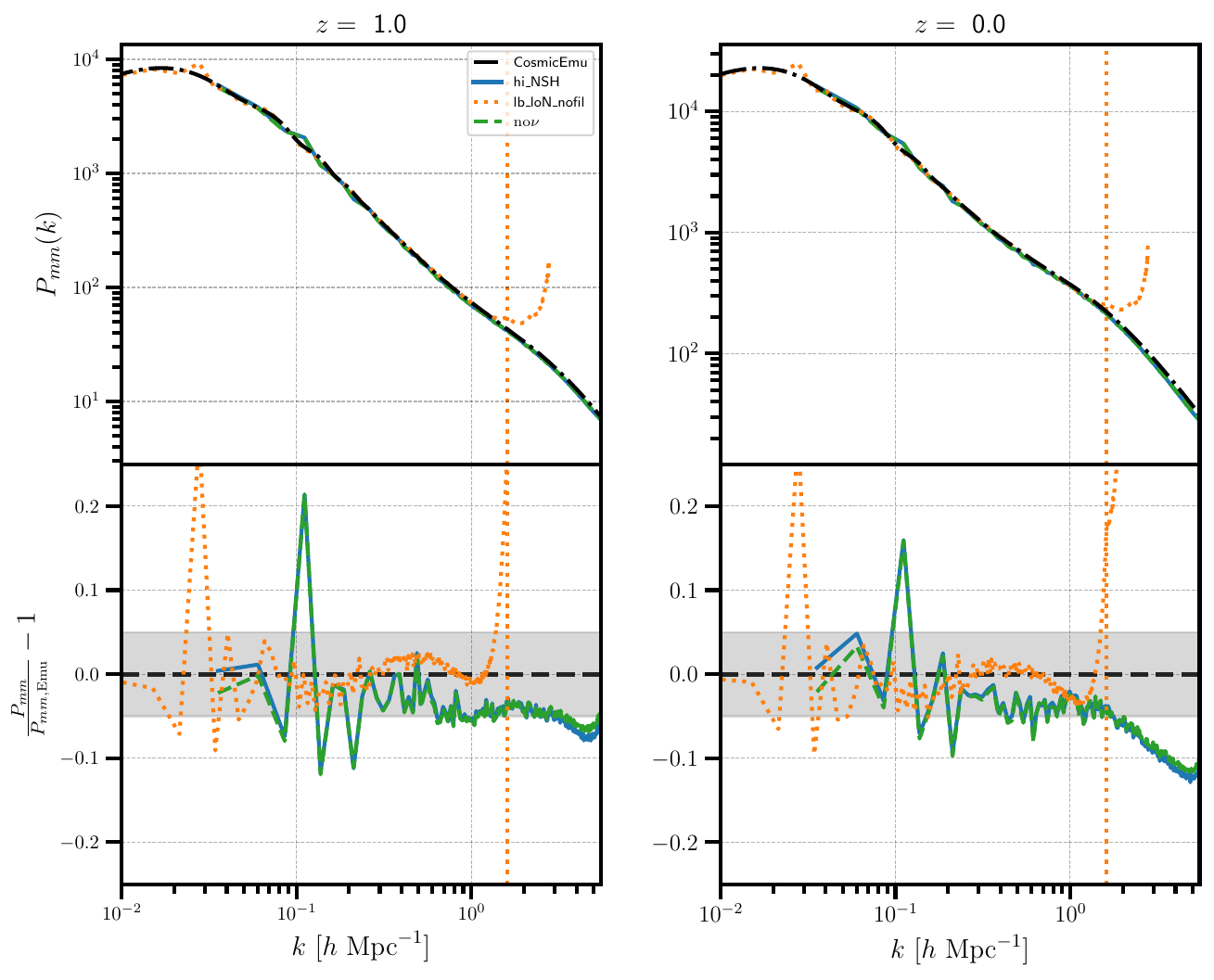}
    \caption{Total matter power spectra in the \texttt{hi\_NSH} ($\nsh=20, \ndi=192$) run [blue solid], the \texttt{lb\_loN\_nofil} ($L=1000~\Mpch$) run [orange dotted], and the 2022 version of the Mira-Titan CosmicEmu \cite{2022arXiv220712345M} [black dash-dotted] at several redshifts.
    We also show the matter power spectrum from a neutrino-less cosmology (``no$\nu$'', with $\Omega_{\nu}=0,~ \Omega_{cdm} = 0.27201$) [green dashed] that has the same numerical parameters as the \texttt{fid\_nofil} run.
    The orange vertical line shows the Nyquist wavenumber for the \texttt{lb\_loN\_nofil} run.
    Lower panels show residuals with shaded bands between $\pm5\%$.
    The residual for the ``no$\nu$'' line is with respect to the CosmicEmu prediction for a neutrino-less cosmology (not shown in the upper panel). See the text for a discussion of the finite-volume effects seen here (the shift between the blue and orange curves).
    }
    \label{fig:emu}
\end{figure}

To make contact between this work and nonlinear growth in total matter power predictions, we compare to the 2022 version of the Mira-Titan Cosmic Emulator \cite{2022arXiv220712345M} in Figure~\ref{fig:emu}.
The simulations used to train the emulator varied the neutrino overdensity parameter $\Omega_{\nu}$, but were not multi-species simulations, so the influence of neutrinos was incorporated only through  1) adding the linear massive neutrino density power spectrum to the cold species power spectrum (as described in Refs.~\cite{2017ApJ...847...50L,2016ApJ...820..108H}), 2) in the computation of $\sigma_{8}$ at $z=0$, and 3) in the rate of homogeneous and isotropic background expansion.
As single-species simulations, these computations neglected scale-dependent growth due to neutrinos in the evolution of the N-body particles, 
so we do not expect our results for the total matter power to exactly agree with the emulator predictions, aside from the errors intrinsic to the construction of the emulator. Nevertheless, at small enough $f_{\nu}$, there should be good agreement. Figure~\ref{fig:emu} shows that this is indeed the case at the percent level (here $f_{\nu}\approx 1\%$ for $\sum m_{\nu} = 0.15~\mathrm{eV}$, where it should be noted that the method used to construct the emulator in the nonlinear regime is effectively an expansion in the leading power of $f_{\nu}$).

The total matter power spectra for the $\nsh=20, \ndi =192$ run at several redshifts approximately match the emulator power spectra at the several percent level.
However, the growth in the emulator prediction is systematically higher than the power in our simulation run for wavenumbers that are low enough that the grid does not impact the power spectrum.
This offset is at approximately the $5\%$ level, which is larger than the $<3\%$ accuracy that would be expected from the emulator test set error \cite{2022arXiv220712345M}.
This effect is not the result of our implementation of the neutrino N-body particle evolution, as a massless neutrino ``no$\nu$'' cosmology shows a similar level of disagreement in Figure~\ref{fig:emu} (green curve).
This disagreement can instead be explained as due to the small box sizes we consider here, as in one of our larger box simulations (orange dotted curve), we see no such $5\%$ offset, and any disagreement between the simulation run and the emulator is at the expected level of $<3\%$ (below the grid scale of that simulation, which is $k\sim \unit{1.0}{\invMpch}$).
These types of finite-volume effects are characterized in more detail in Section 3.3 of Ref.~\cite{2014ApJ...780..111H}, and can be understood more broadly as a combination of a super-sample mode effect \cite{2012JCAP...04..019D,2013PhRvD..87l3504T,2014PhRvD..90j3530L,2014MNRAS.445.3382M,2022arXiv221015647B}, as the small-box simulation is effectively missing large-scale power that is present in the larger simulation box (see e.g. Fig.~6 of Ref.~\cite{2014PhRvD..90j3530L}) and increased variance in individual realizations at smaller box sizes. The estimated variation from the results of  Ref.~\cite{2014ApJ...780..111H} is consistent with that observed in the comparison. In conclusion, Fig.~\ref{fig:emu} shows that the emulator results are consistent with the direct simulations as performed here within the level of errors that are characteristic of both methods.

\subsection{Examination of Neutrino Growth}
\label{subsec:linear}

\begin{figure}[t!]
    \center
    \includegraphics[width=\columnwidth]{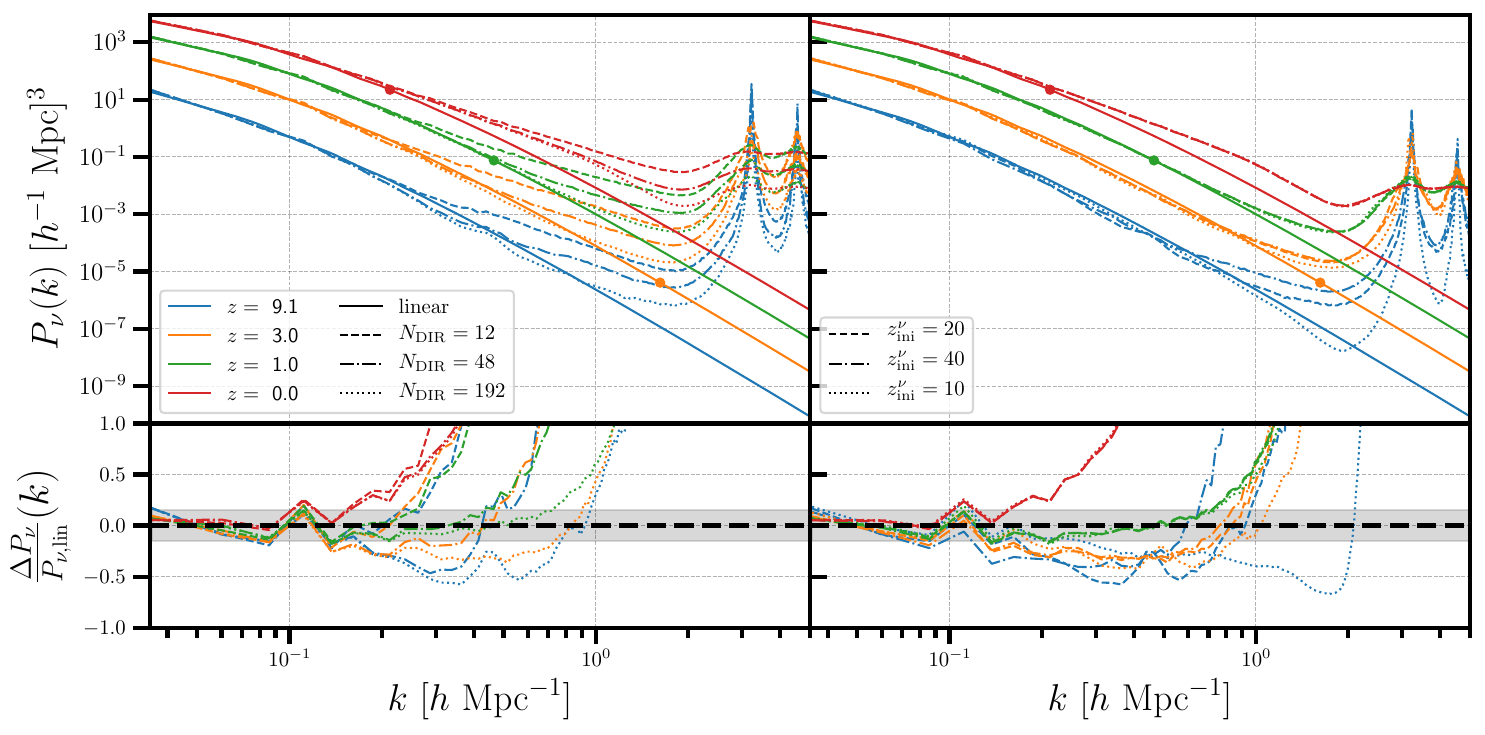}
    \caption{Neutrino power spectra from several simulation runs at several redshifts (colors) compared to linear theory computed using our backscaled transfer functions (solid). 
    \textit{Left:} Neutrino power for the fiducial neutrino initialization redshift ($z_{\rm ini}^\nu = 20$), with a varying number of initial velocity directions: $\ndi = 12$ (dashed, \texttt{fid}), $\ndi = 48$ (dash-dotted, \texttt{mid\_NDIR}), and $\ndi = 192$ (dotted, \texttt{hi\_NDIR}). \textit{Right:} Neutrino power for the highest number of initial velocity directions ($\ndi = 192$) at several choices of starting redshift for the neutrino particles: $z_{\rm ini}^\nu=20$ (dashed, \texttt{hi\_NDIR}), $z_{\rm ini}^\nu=40$ (dash-dotted,, \texttt{hi\_z}), $z_{\rm ini}^\nu=10$ (dotted,\texttt{lo\_z}). At each redshift, we show using a circle the non-linear scale, $k_{\rm nl}$, defined to be the largest scale for which the dimensionless total matter power spectrum, $\Delta_m^2(z) \equiv k^3P_m(k)/2\pi^2$, exceeds unity.
    Lower panels show residuals with respect to (backscaled) linear theory neutrino power spectra and gray bands are between $\pm15\%$.
    }
    \label{fig:linear}
\end{figure}

We complete this section with a closer look at how well the simulated neutrino growth compares to linear theory expectations. In Figure~\ref{fig:linear}, we compare the linear neutrino power spectra at various redshifts to the outputs from the \texttt{fid}, \texttt{med\_NDIR}, and \texttt{hi\_NDIR} runs (previously shown in Figure \ref{fig:tiling_num}). As discussed earlier, the neutrino power spectrum on scales approaching $k_{\rm Nyq}^\nu$ is very sensitive to the number of direction vectors. Ref.~\cite{brandbyge/etal:2019} attributed this to a coherent sampling of the gravitational potential causing a spurious power generation that diminishes as the number of direction vectors increases.\footnote{Ref.~\cite{brandbyge/etal:2019} also found that this spurious power is more significant for higher momentum shells.
By inspecting $q-$dependent power spectra, we also find this to be the case, and that the higher-$q$ shells are more sensitive to the number of directions $\ndi$.}
At high redshift, we would expect that the simulated power spectra numerically converge to linear theory as the number of direction vectors increases, and that this happens most readily on larger scales where the spurious power is weakest. 
However, while we do observe clear numerical convergence as $\ndi$ is increased, the converged result is noticeably {\em below} linear theory for $k \gtrsim \unit{0.1}{\invMpch}$. This systematic suppression is clear at $z = 9$ and persists until late times when non-linear growth eventually dominates the signal. 
While this discrepancy is relatively minor (the simulation with $\ndi=192$ has neutrino power that is 15\% smaller than linear theory), and therefore unlikely to significantly impact the total matter field, it is worth considering any potential systematic errors that could contribute to incorrect neutrino growth. 

There are a number of approximations made in the simulations presented here. In the first place, the momentum-integrated neutrino transfer function is used to set the initial displacement and non-thermal velocities for all momentum shells. In principle, this could be made more accurate by initializing each shell with its own momentum-dependent transfer function. However, given that we do not observe a significant dependence on $\nsh$ seen in Figure~\ref{fig:tiling_sh}, we do not expect this to be a large source of bias in the neutrino power spectrum. Another potential issue is that high thermal motion at early times enables neutrino particles to quickly traverse large distances from their starting point. This behavior may smooth out the power spectrum initially seeded in the displacement field as the neutrinos dynamically readjust to their rapidly changing environments. In this case, we would expect the smoothing to be enhanced with earlier starting times when the thermal motion is more extreme. Another possible bias stems from the non-relativistic treatment of neutrino particles in the simulation; an approximation that becomes increasingly inaccurate at higher redshift and conflicts with the relativistic treatment used in the linear theory reference curves. Finally, the neutrino initial conditions are set using the ZA which is known to be insufficient for cold matter at the neutrino starting redshifts used here. It is unclear if the ZA is also problematic for neutrinos at these redshifts despite their lower density amplitude, but it has been shown in Ref.~\citep{michaux/etal:2021} that systematic power suppression occurs in cold matter simulations with late ZA starts. 

We test the latter three possibilities more closely by running three simulations with different neutrino initial redshifts. These are shown in the right panel of Figure~\ref{fig:linear} where the \texttt{lo\_z}, \texttt{hi\_NDIR}, and \texttt{hi\_z} runs differ only by their use of $z_{\rm ini}^\nu = 10$, 20, and 40, respectively. At $z = 9$, all three runs sit systematically below linear theory\footnote{The linear theory curves are calculated from the backscaling of the $z_{\rm ini}^\nu = 20$ run. There are minor differences in the backscaling for the different $z_{\rm ini}^\nu$ (see Figure~\ref{fig:omegagrav}) but these are small in comparison to the spread in the simulation curves seen here.} at $k \sim \unit{0.2}{\invMpch}$, with this suppression persisting until non-linear growth dominates at $z = 1$. We do observe a small dependence on $z_{\rm ini}^\nu$ at high redshift, with the \texttt{lo\_z} run exhibiting tighter agreement with linear theory compared to the other two simulations. However, this trend is quickly erased by $z = 3$ at which point all three runs show strong agreement on all but the smallest scales. On scales approaching $k_{\rm Nyq}^\nu$, we still see a clear trend with $z_{\rm ini}^\nu$ which we attribute to the earlier start times allowing a greater accumulation of spurious power due to the angular discretization of the tiling method. Hence, the systematic suppression in power does show minor dependence on the neutrino initial redshift, but this is only evident at early times.

More careful tests would be required to determine the exact cause of the neutrino power suppression seen on intermediate scales. Furthermore, it is not clear if this suppression is specific to the tiling scheme or would also be observed in a simulation using the random draw strategy (assuming a sufficient number of particles are used to resolve the neutrino power below the shot noise level). We note that analogous findings have also been observed in mixed cold dark matter plus baryon simulations where the cold dark matter (baryons) grow systematically fast (slow) relative to linear theory expectations when the force resolution is below the mean inter-particle separation \cite{angulo/etal:2013,bird/etal:2020,hahn/etal:2021}. It is conceivable that the same issue manifests in neutrino simulations and may even be exacerbated when the neutrino particle grid is made coarser than that of the cold matter. Even though the discrepancy with respect to linear theory is relatively minor, it is still worth investigating this topic in the future so that systematics in neutrino simulations are properly identified and resolved. 

\section{Conclusions}
\label{sec:conclusions}

A standard approach for initializing cosmological simulations is to backscale the final redshift transfer functions to the initial redshift in a manner that is consistent with the Newtonian forward model of the simulation. This procedure becomes complicated in massive neutrino cosmologies since the simulated growth function acquires a scale dependence that is not easily calculable. We have presented an improved backscaling method for massive neutrino cosmologies that uses an iterative procedure to converge on a self-consistent initialization strategy that preserves the relative neutrino growth predicted by Boltzmann solvers. Our method exploits the known asymptotic limits that the energy density contributing to growth on large and small scales is set by $\Omega_{cb\nu}$ and $\Omega_{cb}$, respectively. When compared to the direct output of Boltzmann solvers at high redshift, our backscaled transfer functions show qualitatively similar behavior to what is seen in massless neutrino cosmologies. This is a notable improvement compared to the common approach where the scale-dependent growth is solved using a two-fluid model that assumes a much different level of neutrino growth compared to Boltzmann solvers. Since our approach maintains the neutrino growth predicted with Boltzmann codes, an added benefit is that we can readily compare the simulated neutrino power spectra with our backscaling model to check that neutrinos are evolving with linear theory expectations.

We applied this backscaling procedure to particle-based neutrino simulations performed using the code \HACC. The neutrino initial conditions follow the tiling framework presented in Ref.~\cite{banerjee/etal:2018} with a number of important modifications. First, we introduce a simple rotation scheme between adjacent momentum shells which reduces spurious power arising from the angular discretization of momentum directions. Second, we omit neutrinos from the short-range gravitational force and apply a sharp truncation to the neutrino density field in Fourier space during the long-range force calculation. This filter can easily be applied in any PM-based force solver and prevents an artificial back-reaction of the initial condition neutrino grid onto the cold matter field. Without any filter, we found perceptible spikes emerge in the cold matter field if the neutrino mass resolution, and the corresponding neutrino shot noise level, is sufficiently coarse.  Since this filter completely removes neutrino power below the particle grid Nyquist frequency, it should only be applied if the particle grid resolution is much finer than the neutrino free-streaming length. More sophisticated treatments would involve applying this filter only at high redshift when both the free-streaming length is larger and the neutrino shot noise level is more comparable to the intrinsic cold matter power spectrum.

We performed a numerical convergence study on the internal parameters of the tiling method. In agreement with Ref.~\cite{brandbyge/etal:2019}, we find that the tiling method is highly sensitive to the number of direction vectors and $\ndi \gtrsim 100$ is required to achieve $\sim10\%$ convergence in the neutrino power spectrum on intermediate scales. The dependence on the number of momentum shells is weaker and is generally converged at the $\sim10\%$ level with $\nsh \gtrsim 10$. In order to prevent the total number of neutrino particles from heavily dominating the total particle count, these findings require a relatively coarse neutrino grid with $N_{cb}/N_\nu \gtrsim 10$. The primary challenge is that the neutrino power spectrum saturates to the shot noise level below the neutrino grid scale with prominent peaks at the grid resonance frequencies. This issue is not unique to the tiling procedure per se, but becomes increasingly problematic as a neutrino grid pushes the noise to larger scales. Hence, while the tiling method effectively removes neutrino shot noise  above the neutrino grid scale, the sacrifice is that care must be taken to ensure that the number of neutrino particles per grid site is large enough to reach numerical convergence and that the gravitational back-reaction from the grid is properly handled. We found that the cold matter power spectrum is largely invariant to changes in the tiling parameters on linear scales. On non-linear scales, however, we found that changes in the tiling parameter choices lead to $\sim 1\%$ random noise.

The simulations presented here converge to a neutrino power spectrum suppressed at the $15\%$ level compared to linear theory for scales $k \gtrsim \unit{0.1}{\invMpch}$ and redshifts $z \gtrsim 1$. We have not determined the exact cause of this discrepancy nor is it clear if this result is unique to the tiling method or would more generally be reproduced in other particle-based methods. Obviously, there are a number of approximations made in the simulation that could preclude accurate comparisons with linear theory. These include the use of the momentum-integrated transfer function for all momentum bins in the initial conditions as well as the fact that we omit all relativistic treatments of neutrinos. It is also possible that high thermal motion at early times erases the initial condition power spectrum and/or that biases seen in mixed dark matter plus baryon simulations extend to the neutrino case. The use of the ZA in initializing the neutrinos at $z_{\rm ini}^\nu = 10-40$ may also be problematic, due to the associated power suppression characteristic of this approximation. Understandably, most studies have focused only on numerical convergence in the total matter field, but it remains important to isolate any errors directly impacting neutrino growth.

We compared the results from our N-body simulations for the total matter power spectrum with the predictions of the Mira-Titan emulator~\cite{2022arXiv220712345M} for a neutrino mass sum of $0.15$~eV, corresponding to $f_{\nu}\approx 1\%$. The results obtained are very encouraging, the agreement being nicely within the estimated errors intrinsic to both methods.

Looking forward, there are several applications where the lessons of this work might be useful.
For example, the methods used for simulating the nonlinear effect of massive neutrinos on probes of LSS can be repurposed to investigate similar effects due to light massive relics (LiMRs, e.g. Ref.~\cite{2022arXiv220307943D}).
Most forecasts of the sensitivity of LSS observations to the presence of LiMRs have been at the level of linear theory or modified perturbation theory \cite{2022PhRvD.105i5029X,2021PhRvD.103b3504D}, though Ref.~\cite{2022MNRAS.516.2038B} recently gave a fully-nonlinear treatment with simulations (using a randomized initialization).
It would be interesting to consider both the reduced computational cost due to shot noise reduction and to explore the relative significance of the numerical artifacts of the tiling initialization in the context of LiMR simulations.
This work also serves to inform how best to include massive neutrinos in a computationally efficient manner in three-species simulations, for example, in conjunction with \CRKHACC \citep{frontiere/etal:2022}.
Future work will more completely characterize to what extent the tiling initialization can be reliably applied to three-species simulations, in particular focusing on the interaction between baryons and neutrinos.

\acknowledgments

We thank Matthew Becker for his role in the early stages of this project. We also thank Arka Banerjee,  Adrian Bayer, Michael Buehlmann, Emanuele Castorina, Joe DeRose, Katrin Heitmann, Patricia Larsen, and Amol Upadhye for helpful conversations. We are grateful to Michael Buehlmann for assistance packaging \SONICC.

This material is based upon work supported by the U.S. Department of Energy, Office of Science, Office of Advanced Scientific Computing Research, Department of Energy Computational Science Graduate Fellowship under Award Number DE-SC0019323. 
This material is further based upon work supported by Laboratory Directed Research and Development (LDRD) funding from Argonne National Laboratory, provided by the Director, Office of Science, of the U.S. Department of Energy under Contract No. DE-AC02-06CH11357.
This research used resources of the National Energy Research Scientific Computing Center (NERSC), a U.S. Department of Energy Office of Science User Facility located at Lawrence Berkeley National Laboratory, operated under Contract No. DE-AC02-05CH11231.
This research has made use of NASA's Astrophysics Data System.
This report was prepared as an account of work sponsored by an agency of the United States Government. Neither the United States Government nor any agency thereof, nor any of their employees, makes any warranty, express or implied, or assumes any legal liability or responsibility for the accuracy, completeness, or usefulness of any information, apparatus, product, or process disclosed, or represents that its use would not infringe privately owned rights. Reference herein to any specific commercial product, process, or service by trade name, trademark, manufacturer, or otherwise does not necessarily constitute or imply its endorsement, recommendation, or favoring by the United States Government or any agency thereof. The views and opinions of authors expressed herein do not necessarily state or reflect those of the United States Government or any agency thereof.

\bibliographystyle{JHEP}
\bibliography{main} 

\appendix

\end{document}